\documentclass[twocolumn, tighten]{aastex62}
\usepackage[utf8x]{inputenc}
\usepackage{amsmath}
\usepackage{amsfonts}
\usepackage{amssymb}
\usepackage{mathrsfs}
\usepackage{bm}
\usepackage[caption=false]{subfig}
\usepackage{ulem} 
\usepackage{esint} 

\defcitealias{1992ApJ...389..129A}{AL92}
\defcitealias{2015ApJ...815L..21F}{FD15}
\defcitealias{2002MNRAS.330..950O}{OL02}
\defcitealias{2015MNRAS.448.3806L}{L15}
\defcitealias{1990ApJ...358..495S}{STAR90}
\defcitealias{2001AJ....121.1776T}{T01}

\begin{document}
\title{Eccentric Modes in Disks With Pressure and Self-Gravity}
\author[0000-0002-5319-3673]{Wing-Kit Lee}
\author[0000-0001-8291-2625]{Adam M. Dempsey}
\author{Yoram Lithwick}
\affiliation{Center for Interdisciplinary Exploration and Research in Astrophysics (CIERA)\\ and
Department of Physics and Astronomy, Northwestern University, 2145 Sheridan Road, Evanston, IL 60208, USA}

\correspondingauthor{Wing-Kit Lee}
\email{wklee@northwestern.edu}
\shorttitle{Eccentric Modes in Disks With Pressure and Self-Gravity}
\shortauthors{LEE et al.}

\begin{abstract}
Accretion disks around stars, or other central massive bodies, can support long-lived, slowly precessing $m=1$ disturbances  in which the fluid motion is nearly Keplerian with non-zero eccentricity.  We study such ``slow modes'' in disks that are subject to both pressure and self-gravity forces. We derive a second-order WKB dispersion relation that describes the dynamics quite accurately, and 
show that the apparently complicated nature of the various modes can be understood in a simple way with the help of a graphical method. We also solve the linearized fluid equations numerically, and show that the results agree with the theory. We find that when self-gravity is weak ($Q\gtrsim 1/h$, where $Q$ is Toomre's parameter, and $h$ is the disk aspect ratio) the modes are pressure dominated. But when self-gravity is strong ($1<Q\lesssim 1/h$),  two kinds of gravity-dominated modes appear: one is an aligned elliptical pattern and the other is a one-armed spiral. In the context of protoplanetary disks, we suggest that if the radial eccentricity profile can be measured, it could be used to determine the total  disk mass.
\end{abstract}


\section{Introduction}

Eccentric distortions exist in many kinds of nearly Keplerian astrophysical disks, 
including planetary rings, protoplanetary disks, and around supermassive black holes in galactic nuclei. In the Solar System, eccentric rings include the Maxwell ringlet of Saturn's C ring \citep[see,][]{1983Sci...222...57E,2014Icar..241..373N,2016Icar..279...62F} and the $\varepsilon$ ring of Uranus \citep{1978AJ.....83.1240N}. 
Some off-centered galactic nuclei \citep[e.g., M\,31,][]{1993AJ....106.1436L} can be explained by an eccentric stellar disk \citep{1995AJ....110..628T}. In the context of star and planet formation, many circumstellar disks are found to have a large-scale asymmetry  \citep[e.g.,][]{2013Sci...340.1199V,2017ApJ...840...32T,2017A&A...607A..55V}.
And an eccentric cavity has recently been observed in a protoplanetary disk for the first time \citep{2018ApJ...860..124D}.

In this paper, we explore how eccentric distortions in a nearly Keplerian disk can be maintained via gas pressure and self-gravity. We focus on long-lived slowly-precessing eccentric $(m=1)$ modes, which are of long wavelength and hence less susceptible to viscous damping. Such modes may be excited by a companion body or the passage of an external object, but we leave the topic of excitation to future work.

Key previous work includes the following: \citet*{1989ApJ...347..959A} and \citet*{1990ApJ...358..495S} (hereafter \citetalias{1990ApJ...358..495S}) found that $m=1$ modes can be unstable if the disk is sufficiently massive, 
 even if the disk is Toomre-stable. \citet{1999MNRAS.308..984L} studied angular momentum transport by a  nonlinear one-armed spiral that has zero-frequency.

\citet[][hereafter \citetalias{2001AJ....121.1776T}]{2001AJ....121.1776T} used WKB theory and numerics to study long-lived slow modes in self-gravitating disks. Although he did not explicitly include pressure, he modeled its effect with a softening
parameter  for self-gravity.

\citet{Pap2002}  studied the spectrum of eccentric modes in two disk models with both gas pressure and self-gravity---and also with and without planets. Although his study was mostly numerical, he  provided physical explanations of his results. And \citet{2016MNRAS.458.3221T} extended upon that work by including a model for a 3D disk, as well as mean-motion resonances with a planet.

Theoretical studies of non-self-gravitating (i.e., pressure-only) disks \citep*{2008MNRAS.388.1372O, 2009MNRAS.400.2090S} reveal that eccentric modes {are} described by a Schr\"{o}dinger-like equation, and that eccentric
 modes can be trapped in a disk in the same way that a quantum particle is trapped by a potential. In a forthcoming paper (Lee, Dempsey, \& Lithwick, \emph{in prep.}, hereafter Paper II), we study the general conditions under which an eccentric mode is  trapped in a pressure-only disk.
A few other studies of {non-self-gravitating} eccentric modes include: small-scale instability using a local shearing-sheet model of an eccentric basic state \citep{2014MNRAS.445.2621O,2014MNRAS.445.2637B}; eccentric magneto-rotational instability \citep{2018ApJ...856...12C}; three-dimensional nonlinear theory \citep{2001MNRAS.325..231O,2018MNRAS.477.1744O}; and nonlinear simulation \citep{2016MNRAS.458.3739B}. 

We return to  a more detailed discussion  of prior results in \S \ref{sec:discussion}.

This paper is organized as follows: In \S \ref{sec:formulation}, we present the linearized equations of motion; in \S \ref{sec:suiteofsixmodels} we present numerical solutions to that equation (eigenmodes and eigenfrequencies) for a fiducial suite of six disk models that have varying relative strength of pressure to self-gravity; in \S \ref{sec:asymptotic_theory} we present the second-order WKB theory; and in \S \ref{sec:wkb2analysis} we apply the WKB theory to explain the numerical results from \S \ref{sec:suiteofsixmodels}. At the end of the paper we briefly consider other disk models before discussing some implications of our results. 

\section{Formulation}
\label{sec:formulation}

\subsection{Equations of motion}

We consider a two-dimensional\footnote{It is not clear whether a purely two-dimensional treatment is adequate \citep{2008MNRAS.388.1372O}. We address this concern in \S \ref{sec:discussion_connection}.} fluid disk orbiting a central star that is subject to both pressure and self-gravity forces. The continuity equation reads
\begin{align}
\label{eq:basic1_1a}
\partial_t \Sigma + \nabla\cdot (\Sigma {\bm u}) &= 0 \ ,
\end{align}
where $\Sigma$ and ${\bm u}$ are the gas surface density and velocity, respectively. The momentum equation is given by
\begin{align}
\label{eq:basic1_1b}
\partial_t {\bm u} + {\bm u} \cdot \nabla {\bm u} &= -\frac{GM_\star}{r^2}-\frac{1}{\Sigma}\nabla P - \nabla \phi \ ,
\end{align}
where $M_\star$ is the stellar mass, $P$ is the two-dimensional pressure and $\phi$ is the gravitational potential due to self-gravity, which is governed by the Poisson equation
\begin{align}
\label{eq:basic1_2}
	\nabla^2 \phi = 4\pi G \Sigma \delta(z) \ ,
\end{align}
where $\delta(z)$ is the Dirac delta function representing the razor-thin disk. We ignore the indirect potential 
  because it does not affect the slow $m=1$ modes (\citetalias{2001AJ....121.1776T}; \citealt{2016MNRAS.458.3221T}; Appendix \ref{sec:linearization}).

\subsection{Basic State}
\label{sec:basic}

The basic state of the disk is time-independent and axisymmetric. The equilibrium azimuthal velocity $r\Omega$ is determined from radial force balance to be
\begin{align}
\label{eq:basic2_1a}
r\Omega^2 = \frac{GM_\star}{r^2} + \frac{1}{\Sigma}\frac{dP}{dr} + \frac{d\phi_{0}}{dr} \ ,
\end{align}
where
$P$ and $\Sigma$ here and henceforth refer to their equilibrium values, and
 $\phi_{0}$ is the resulting equilibrium potential.

\subsection{Linearized equations of motion}
\label{sec:slowmodeapprox}

We consider   {\it slow} modes, i.e., normal modes whose frequency is less than
the orbital frequency everywhere in the disk: $|\omega|\ll \Omega$. In other words, slow modes have corotation lying outside of the disk. The linearized equation of motion can be expressed simply in terms of the complex eccentricity \citep{Pap2002}
\begin{align}
\label{eq:basic3_1}
E = |E| e^{-i\varpi} \ ,
\end{align}
where the real eccentricity $|E|$ and longitude of  pericenter $\varpi$ are both functions of radius. The governing equation of the disk eccentricity has been derived in the literature by previous authors \citep[e.g.,][]{Pap2002,2006MNRAS.368.1123G,2016MNRAS.458.3221T}. We also provide a self-contained derivation in Appendix \ref{sec:linearization}. After replacing $\partial/\partial t\rightarrow -i\omega$, and taking the perturbation to be adiabatic with a 2D adiabatic index $\gamma$,  the linearized equation reads 
\begin{align}
\nonumber
\omega E &= \underbrace{
{1\over 2r^3\Omega\Sigma}\left[\frac{d}{dr}\left(\gamma r^3 P \frac{dE}{dr}\right) + r^2 \frac{dP}{dr} E\right]
}_\text{pressure} \\
\label{eq:basic3_2}
& \underbrace{-
{1\over 2r^3\Omega}
\left[
 r\frac{d}{dr}\left(r^2\frac{d\phi_0}{dr}\right)E+  \frac{d}{dr}(r^2\phi_1)
\right]
 }_\text{self-gravity} \ ,
\end{align}
where $\Omega$ here and henceforth refers to the Keplerian value ($\sqrt{GM_*/r^3}$), and $\phi_1$ is the self-gravity potential for the perturbation ($m=1$).
  Explicit equations for 
  $\phi_0$ and $\phi_1$  are 
given in Equation   \eqref{eq:app1_21}, which expresses the two potentials 
   as integral transforms of  $\Sigma$ and $\Sigma_1$, respectively, thus yielding a closed system of equations (after replacing $\Sigma_1\rightarrow E$ via Equation \eqref{eq:app1_6a}).

Equation \eqref{eq:basic3_2} is a linear integro-differential equation  that we will
solve for eigenvalues
$\omega$ and  eigenfunctions $E(r)$.
 Before doing so,  the relative importance of pressure and self-gravity can be assessed
  by replacing $dr\rightarrow r$ in derivatives and integrals. One finds, 
after defining the dimensionless quantities
\begin{eqnarray}
h = \frac{c}{r\Omega} \quad \text{and} \quad \mu = \frac{\pi G\Sigma}{r\Omega^2} \ , \label{eq:hmu}
\end{eqnarray}
where the former is the aspect ratio of the disk and the latter is the ratio of disk mass to stellar
mass (within order-unity factors), and
\begin{equation}
c=\sqrt{\gamma P/\Sigma}
\end{equation}
is the sound-speed, that the
pressure and self-gravity terms become
\begin{eqnarray}
{\rm pressure} &\sim& h^2\Omega E, \\
\text{self-gravity} &\sim& \mu\Omega E \ .
\end{eqnarray}
In other words, pressure causes mode periods to be longer than the orbital time
by $\sim 1/h^2$, and self-gravity causes them to be longer by $\sim 1/\mu$. 
We also introduce a third dimensionless quantity that characterizes the relative
strengths of self-gravity and pressure, 
\begin{equation}
g= {\mu\over h^2}= {\pi G\Sigma r\over c^2} \ ,
\label{eq:gdef}
\end{equation}
which is 
similar to
\citeauthor[][]{1999MNRAS.308..984L}'s (\citeyear{1999MNRAS.308..984L})
$\sigma^2$. The $g$ parameter differs from Toomre's $Q$  that characterizes axisymmetric collapse, 
$Q\sim h/\mu$.
   Therefore a disk may support self-gravity-dominated slow modes ($g\gtrsim 1$), yet
be stable to axisymmetric collapse ($Q\gtrsim 1$), provided 
\begin{equation}
h^2\lesssim \mu\lesssim h \ .
\end{equation}

\section{Fiducial Suite of Six Models: Numerical Simulations}
\label{sec:suiteofsixmodels}

In this section we numerically solve Equation \eqref{eq:basic3_2} for the eigenfrequencies and eigenmodes in a suite of six disk models. In the subsequent two sections, we explain the numerical results with a WKB theory.

\begin{figure}
	\centering
	\includegraphics[trim={0.7cm 0.6cm 0 0},clip,width=0.47\textwidth]
	{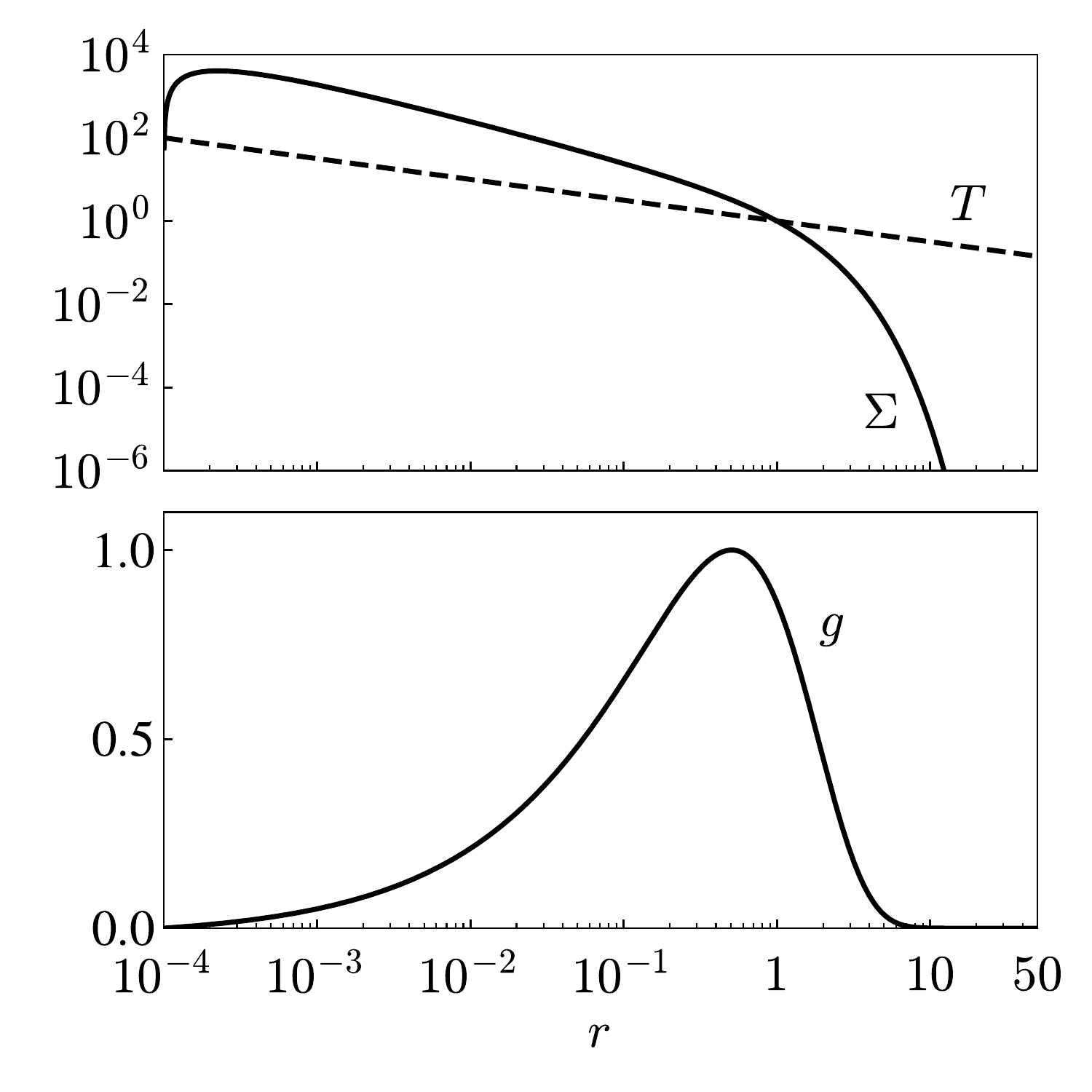}
	\caption{Unnormalized background profiles used in the suite of six models.}
		\label{fig:basicstate}
\end{figure}

\subsection{Background Profiles}
\label{sec:bm}
Our choice of disk profile (Figure \ref{fig:basicstate}) is guided by the structure of protoplanetary disks. For the  temperature profile, we choose  a power-law
\begin{equation}
\label{eq:background_T}
T(r)=C_T \, r^{-q} \ ,
\end{equation}
 with constants $C_T$ and $q$.
For the surface density, we choose a power-law with  exponential cutoff at $r\gtrsim 1$, and  an inner tapering function near $ r_{\rm in}=10^{-4}$:
\begin{equation}
\label{eq:background_density}
\Sigma(r) =C_\Sigma f_{\rm tap}(r)r^{-p}\exp({-r^{2-p}}) \ ,
\end{equation}
where the tapering function is $f_{\rm tap}= 1-\sqrt{r_{\rm in}/r}$, which vanishes at $r_{\rm in}$ and reaches unity at 
  $r\gtrsim$ a few$\times r_{\rm in}$.
The factor inside the exponential ($r^{2-p}$) follows from the 
self-similar solution of a viscously evolving accretion disk  that has a power-law viscosity \citep{1974MNRAS.168..603L}. 
The background pressure is
\begin{equation}
P=\Sigma\,T \ .
\end{equation}
We choose  the parameter values
\begin{equation}
\gamma=3/2 \ , \quad q=1/2 \ , \quad \text{and} \quad p=1 \ , \label{eq:paramvals}
\end{equation} 
where this $\gamma$ is the 2D adiabatic index that corresponds to the 3D index $\gamma_{3D}=5/3$ \citep[e.g.,][]{2001ApJ...553..174G}. Also shown in Figure \ref{fig:basicstate} is the unnormalized profile of
$g$ (Equation \eqref{eq:gdef}), showing that, for our chosen parameters, self-gravity is most important relative to pressure near the disk's outer cutoff.

It remains to specify $C_\Sigma/C_T$ or, 
 equivalently, the
maximum of the $g(r)$ profile. We investigate six values:
$g_{\rm max}=0.5, 1, 2, 5, 10$, and 20. Note that
 it is only the ratio $C_\Sigma/C_T$ that is needed, rather than  $C_\Sigma$ and $C_T$
separately because, 
from Equation \eqref{eq:basic3_2},
 the pressure terms are $\propto C_T$ and the self-gravity terms are $\propto C_\Sigma$. Hence 
  only the ratio of constants appears in the equation once  we normalize $\omega$ by $C_T$. 
  We  choose to normalize $\omega$ by
\begin{equation}
\omega_{0}= C_T {\gamma \over 2(\Omega r^2)\vert_{r=1}}
={\gamma\over 2}\left( {T\over\Omega r^2} \right)_{r=1}= {1\over 2}(\Omega h^2)_{r=1} \ ,
\end{equation}
where, in addition to $C_T$, we  absorb the normalization of $\Omega r^2$ into $\omega_{0}$ to
give it the dimensions of frequency. 
With this normalization, when solving Equation \eqref{eq:basic3_2} we may
(i) set $\Omega=r^{-3/2}$, because its coefficient has been absorbed into $\omega_{0}$;
(ii)  set   $C_T=1$, because its normalization is irrelevant; 
and (iii) set
$C_\Sigma$ via the relation $g_{\rm max}=\pi G\Sigma / (\gamma T)$ evaluated at the radius
where $g$ reaches  its maximum.  That  latter step yields a number for the product $GC_\Sigma$, which is what enters into the inversion of Poisson's equation for $\phi_0$ and $\phi_1$.

 \subsection{Numerical Method}
 \label{sec:full_methodofsolution}
 
We solve Equation \eqref{eq:basic3_2} for the eigenfunctions and eigenfrequencies with a finite difference matrix method. Since our numerical implementation is mostly standard, we relegate details to Appendix \ref{sec:methodofsolution}.  However, we highlight here one notable feature. For the two-dimensional problem, the Poisson inversion involves a kernel whose diagonal element is formally infinite. We remove that infinity  by integrating  analytically across the dangerous grid element, adapting \citet[][]{1996ApJ...460..855L} to the case of $m=1$ slow modes (similar to, but different from, T01's approach.) Our method is  both simple to implement and has a faster convergence rate than methods
that rely on a  softening parameter (Appendix \ref{sec:convergence}). In addition, although we do not make use of softening in this paper (except in the appendix, for comparison purposes),  we also provide in Appendix \ref{sec:methodofsolution} a  more efficient way to invert Poisson's equation with softening, improving upon the treatments of \citetalias{2001AJ....121.1776T} and \citet{2016MNRAS.458.3221T}.

All of the numerical solutions in the body of this paper are  run on a grid of 2048 points that are logarithmically spaced between the boundaries at $r=1.001r_{\rm in}$ and $r=50$. For boundary conditions, we set the Lagrangian pressure perturbation to zero at the disk boundaries, which is equivalent to setting  $dE/dr=0$ there
\citep[][and Appendix \ref{sec:linearization}]{Pap2002}. The modes found in this paper are not affected by boundary conditions because their amplitudes are concentrated away from the edges. Or, to be more precise, although we do find some modes with $|E|$ concentrated near the boundary, as we show below  the relevant amplitude is $(r^3P)^{1/2}E$, and that quantity is always concentrated away from the boundaries (See also Paper II). 

\subsection{Numerical Results}
\label{sec:nr}
 
\begin{figure*}
\centering
\includegraphics[trim={0.7cm 1.8cm 0cm 0cm},clip,width=\textwidth]{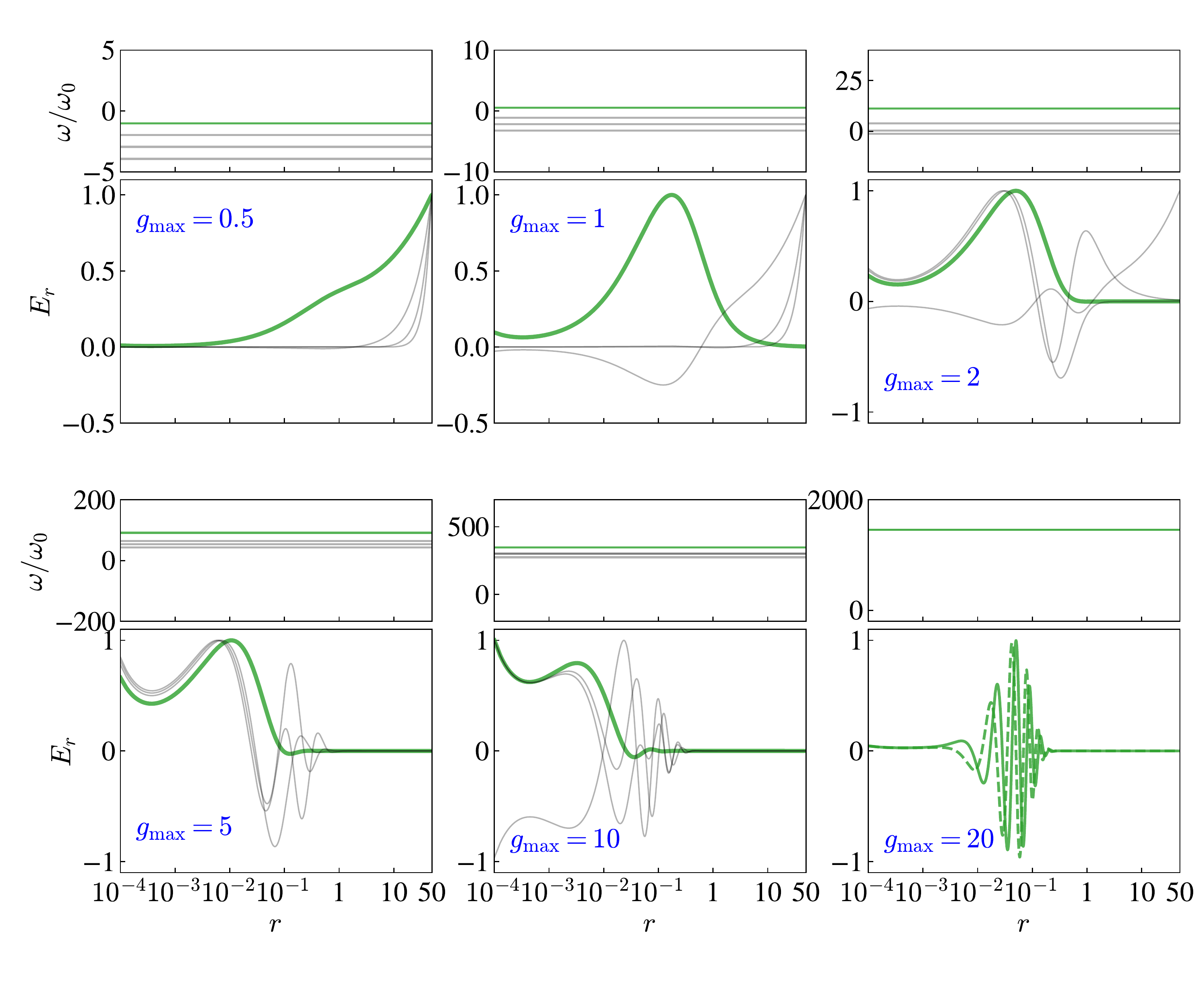}
\caption{
Numerical eigenvalues ($\omega/\omega_0$) and eigenfunctions (real part of $E(r)$) in the suite of six models, with the fundamental 
mode colored green. The eigenvalues are displayed simply as horizontal lines, in anticipation of Figure \ref{fig:Collection} below in which we compare numerical results with theory. The eigenfunctions are all real and have arbitrary amplitudes in the linear theory. Each of the panels with $g_{\rm max}<20$ also shows the three modes with the next smallest number of nodes. For $g_{\rm max}=20$, we only show the fundamental modes, of which there are two (shown as solid and dashed)
with identical frequency; in other words, this eigenvalue is doubly-degenerate.}
\label{fig:eigensolutions}
\end{figure*}

Figure \ref{fig:eigensolutions} shows the numerical eigenfunctions and eigenvalues.  Only  the modes with the fewest numbers of nodes are shown because those have the longest wavelengths, and hence are 
 least susceptible to viscous damping. They are also the ones 
  most likely to be observable if they exist.  The fundamental mode, i.e., the mode with
  fewest number of nodes, is shown in green.
 There are some notable features that we seek to explain in the following sections: 
   (i) as $g_{\rm max}$ is increased, the fundamental mode has increasingly small wavelength;
     (ii)  the fundamental mode has the highest (most positive) frequency, while higher harmonics have lower
     frequencies; (iii) for $g_{\rm max}=0.5$, the fundamental mode is retrograde ($\omega<0$), while for higher $g_{\rm max}$ it is prograde; and (iv) the eigenfunction is concentrated at 
     $r\gg 1$ for the smallest $g_{\rm max}$, and it shifts to smaller $r$ with increasing 
     $g_{\rm max}$. 
     
\section{WKB Theory}
\label{sec:asymptotic_theory}

\subsection{Second-Order WKB Dispersion Relation (WKB2)}
\label{sec:asymptotic_dr2}

To make sense of the numerical results, we adopt the WKB (or tight-winding)
approximation  by assuming
 that the radial wavelength
of the disturbance is  smaller than the lengthscale of the background.
Setting $d/dr\rightarrow ik$ in   Equation \eqref{eq:basic3_2}\footnote{
With our sign convention, modes with $k>0$ are trailing and ones with $k<0$ are leading. 
Furthermore, modes with $\omega>0$ are prograde, and ones with $\omega<0$ are retrograde.
}, 
 and taking the limit  $|kr|\gg 1$, the terms on the right-hand side 
that scale with the highest powers of $k$ are 
\begin{eqnarray}
\nonumber
 -k^2 c^2 E -\frac{ik r^2}{2r^3 \Omega} {2\pi G  ikr \Sigma\over |k|}  E \ ,
\end{eqnarray}
where the second term follows from inverting Poisson's equation for $\phi_1$, as
shown in Equation (\ref{eq:app5_6b}), and  then inserting Equation \eqref{eq:app1_6a} for $\Sigma_1$. 
Simplifying,  the  dispersion relation is
\begin{eqnarray}
\omega=-{k^2c^2\over 2\Omega}+{\pi G\Sigma\over\Omega}|k|  \ \ \ {\rm  [WKB0]} \ .
\label{eq:lo}
\end{eqnarray}
 In a slight abuse of nomenclature, we call this the zeroth-order dispersion relation (or WKB0 for short).
Whereas it might seem natural to say that the second term is  higher order than the first because it
 scales with a lower power of $|k|$,  
 the second term can be larger than the first for $G\Sigma/c^2$ sufficiently large, while still
 working in the limit $|kr|\gg 1$. Therefore we group the two terms together at zeroth order.

Unfortunately, WKB0 is insufficiently accurate for our purposes because the fundamental 
mode can vary on the scale of the background, $|kr|\sim 1$ (Figure \ref{fig:eigensolutions}). 
Motivated by studies of galactic spiral waves, we therefore extend WKB0
 by including terms in the dispersion relation that are smaller by factors of $\sim 1/|kr|^2$ \citep{1978ApJ...226..508L,1989ApJ...338..104B}.  
As we demonstrate below, this second-order dispersion relation (``WKB2'') turns out to be surprisingly accurate---even when applied to modes with $|kr|\sim 1$.

We derive WKB2 in Appendix \ref{sec:derivationofDR2}. The result is
 \begin{align}
\label{eq:asymp1_1a}
\omega = \underbrace{\left(-\frac{k^2 c^2}{2\Omega}+\omega_p\right)}_\textrm{pressure} + \underbrace{\left(\frac{\pi G \Sigma}{\Omega}|k|+\omega_g\right)}_\textrm{self-gravity} \quad \text{[WKB2]} \ ,
\end{align}
where the second-order corrections are
\begin{align}
\label{eq:asymp1_1b}
\omega_p &= -\frac{c^2}{2\Omega}\left[(r^3 P)^{-1/2}\frac{d^2}{dr^2}(r^3 P)^{1/2}-\frac{1}{\gamma r P}\frac{dP}{dr}\right],
\end{align}
and
\begin{align}
\nonumber
\omega_g&= - \frac{1}{2\Omega r^2}\frac{d}{dr}\left(r^2\frac{d\phi_0}{dr}\right)& \\
-& \left(\frac{\pi G \Sigma}{\Omega r}\right)\frac{3}{2\sqrt{k^2 r^2 + 9/4}}
\left\lbrace\frac{d}{d\ln r}\ln\left[\frac{\Sigma}{(r^3 P)^{1/2}}\right]-\frac{1}{4}\right\rbrace.
\label{eq:omg}
\end{align}
Two non-trivial steps are used in the appendix to derive $\omega_p$ and $\omega_g$.
First, for $\omega_p$,
  Equation \eqref{eq:basic3_2} is cast into its ``normal form'' \citep[e.g.,][]{2007AN....328..273G} by employing an integrating factor, i.e., by changing variables from $E$ to $(r^3P)^{1/2}E$, which
  ensures there are no purely first derivative
  terms in the  pressure contribution to Equation  \eqref{eq:basic3_2}. (See also Paper II).
And second, in deriving $\omega_g$, Poisson's equation for $\phi_1$ is inverted by   keeping not only the leading WKB term, but also corrections that are smaller by $\sim 1/|kr|^2$ \citep{1979SJAM...36..407B}.

For future convenience, we 
write WKB2 in a simpler form by introducing   the dimensionless wavenumber
\begin{equation}
K\equiv kr \ ,
\end{equation}
in which case WKB2 becomes
 \begin{eqnarray}
 \label{eq:wkb2}
 \omega = -{1\over 2}K^2h^2\Omega+\mu |K|\Omega+\omega_p+\omega_g  \ .
 \label{eq:om}
\end{eqnarray}
Furthermore, $\omega_p$ and $\omega_g$  have simple expressions  when $h$ and $\mu$ are power-laws:
\begin{eqnarray}
\omega_p&=&C_p h^2\Omega \ , \\
\omega_g&=&\left( C_{g1}+{C_{g2}\over \sqrt{1+4 K^2/9}} \right)\mu\Omega \ ,
\end{eqnarray}
where the $C$s are order-unity constants.  For example, for our fiducial parameters (Equation \eqref{eq:paramvals}), $C_{p}=-0.6$, $C_{g1}=-1.0$ and $C_{g2}=2.0$ at $r\lesssim 1$ where the power-law applies.

Note that $\omega_p$ is smaller than the leading pressure term by $\sim 1/|K|^2$, and the two terms within $\omega_g$ are smaller than the leading self-gravity term by $\sim 1/|K|$ and $\sim 1/|K|^2$, respectively.

\subsubsection{Comparison with   \cite{1964ApJ...140..646L}}
 
Both WKB0 and WKB2 differ  from the standard dispersion relation  derived in the galactic context \citep{1964ApJ...140..646L,1966PNAS...55..229L}: $(\omega_{\rm LinShu}-m\Omega)^2=\kappa^2-2\pi G\Sigma|k|+k^2c^2$.
After setting $m=1$, specializing to slow modes ($|\omega|\ll \Omega$),
and approximating $(\Omega^2-\kappa^2)/(2\Omega) \approx \Omega-\kappa$ as 
is true for a nearly point-mass potential, we arrive at
\begin{equation}
\omega_{\rm LinShu}\approx -{k^2c^2\over 2\Omega}+{\pi G \Sigma\over \Omega}|k|+(\Omega-\kappa)	\label{eq:linshu}
	\end{equation}
(\citealt{1999MNRAS.308..984L}; \citetalias{2001AJ....121.1776T}; \citealt{Pap2002}).
The explicit
  form for the third term,  $\Omega-\kappa$, is given in Equation \eqref{eq:app1_11}.  
 We see that the first two terms in $\omega_{\rm LinShu}$ are the same as WKB0, but
 the final term differs from the correct one, $\omega_p+\omega_g$. 
    There are two errors in Equation (\ref{eq:linshu}):
 (i) the pressure contribution to $\Omega-\kappa$ differs from $\omega_p$; and (ii)  the gravity contribution to $\Omega-\kappa$ is equal only to the first term in $\omega_g$, while
 not accounting for the second ($k$-dependent) one. 
  This problem with the Lin-Shu dispersion relation is only important for slow modes, because fast modes
    have a $k$-independent term at leading order, and so the higher-order $k$-independent corrections are subdominant.
 
 \subsection{DRM and Mode Types  (WKB1.5)}
 \label{sec:mp}
 
  We seek to decipher the
  properties of  modes encoded by WKB2. 
But the analysis  is complicated  by the $k$-dependence of $\omega_g$.  
 Therefore in this and the next subsection we
  simply ignore that dependence  by setting $k=0$ in $\omega_g$, an 
  approximation we call WKB1.5 
   because
it is correct to second order for pressure, but only to first order for self-gravity. WKB1.5 is simpler to analyze because
it is similar to the Lin-Shu dispersion relation, which is well-understood
  \citep[e.g.,][]{2008gady.book.....B}. And, as with Lin-Shu, WKB1.5
 has only two roots for $|k|$,
 whereas WKB2 has an extra ``very-long'' branch \citep{1978ApJ...226..508L,1989ApJ...338..104B}.  
As we show below,   analysis of WKB2 will be a straightforward extension of that of WKB1.5.

The two branches of WKB1.5 are  simplest to see when $|K|$ is large. 
Dropping $\omega_p+\omega_g$ in  Equation (\ref{eq:om}), one sees  
that pressure dominates at high-$|K|$ and self-gravity at low-$|K|$, with the transitional
$|K|\sim \mu/h^2=g$ 
 (Equation \ref{eq:gdef}), i.e., 
\begin{eqnarray}
\label{eq:wkb15_1a}
|K| \gtrsim g  &\Rightarrow& \mbox{pressure-dominated (``short-branch'')} \\
\label{eq:wkb15_1b}
|K| \lesssim g  &\Rightarrow&\mbox{self-gravity-dominated  (``long-branch'')} \ .
\end{eqnarray}
\citep{2008gady.book.....B}.

\begin{figure*}[t!]
	\centering
	\includegraphics[trim={1cm 0.5cm 2cm 0.5cm},clip,width=\textwidth]{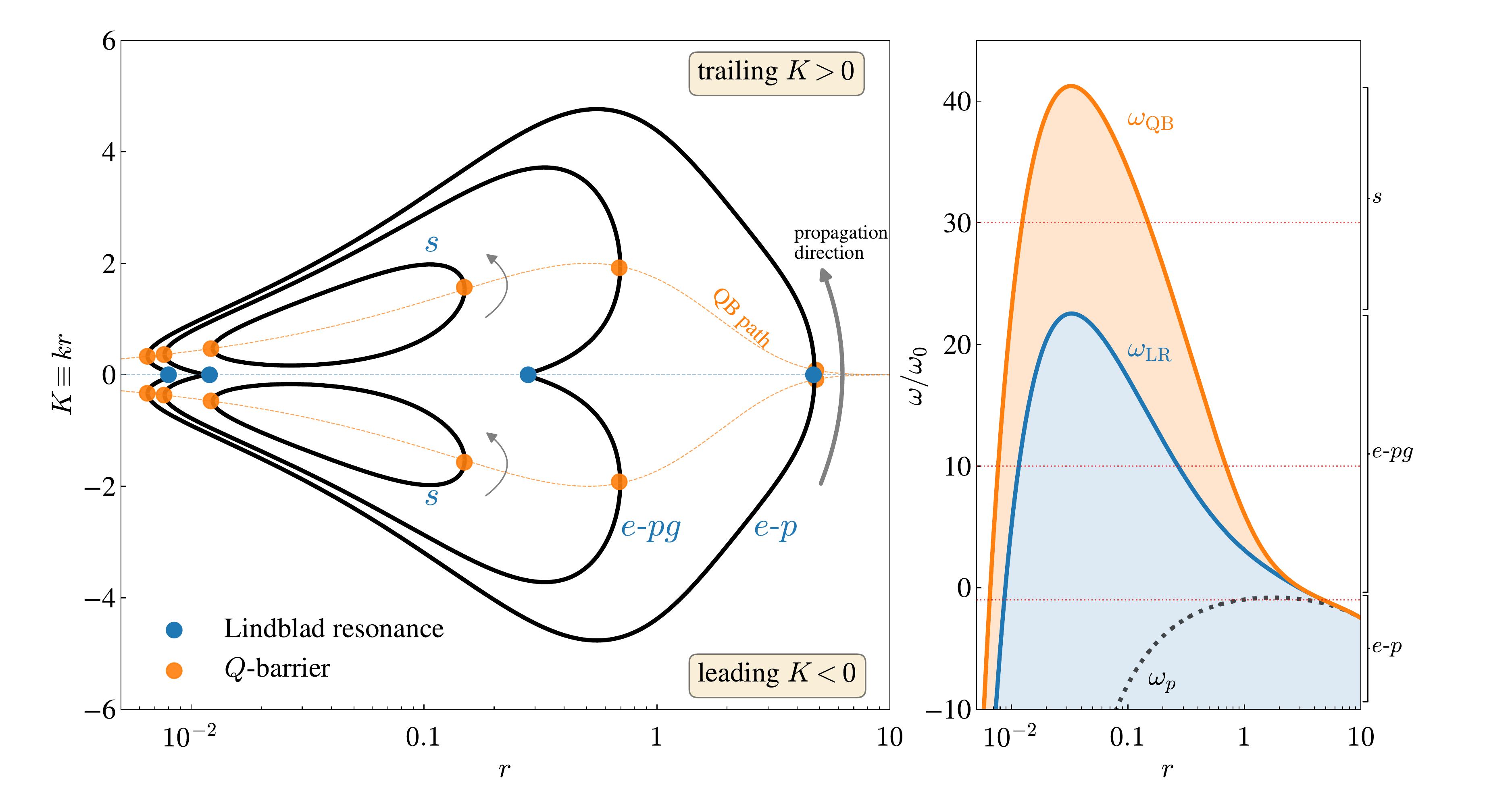}
	\caption{({\emph{Left}}) The Dispersion Relation Map (DRM), showing contours of constant $\omega$ (black solid). For this plot, the dispersion relation is given by Equations \eqref{eq:asymp1_1a}$-$\eqref{eq:omg}, after setting  $k=0$
	 in the expression for $\omega_g$, i.e., the WKB1.5 approximation. The background profiles are taken from the $g_{\rm max}=2$ model  from \S \ref{sec:suiteofsixmodels}. Also shown in this plot are the LR path (blue dotted), and the QB path (orange dotted). 
	 ({\emph{Right}}) Frequency level diagram, showing the values of $\omega$ along the QB and LR paths in the DRM. The three horizontal red lines correspond to the frequencies of the  contours in the {left} panel.
	  }
	\label{fig:illustration}
\end{figure*}

For a global view,
we plot contours of constant $\omega$ in the $K$-$r$ plane, which we call a ``dispersion relation map,'' or DRM. Similar plots appear in the literature (e.g., \citealt{1970ApJ...160...99S}; \citealt{1989ApJ...338..104B}; \citetalias{2001AJ....121.1776T}), but they are perhaps insufficiently common given their usefulness. Figure \ref{fig:illustration} (left panel) presents a DRM in which the background profiles are those of the $g_{\rm max}=2$ fiducial model. Contours are shown at three different values of $\omega$, with each closed contour representing a possible mode. 

A DRM's structure is governed by its turning points, 
which is where the contours reach extrema in $r$.
There are two kinds of turning points in WKB1.5 \citep{1992pavi.book.....S}:
\begin{enumerate}
\item {\it Q-barrier (QB)}. At a QB the group velocity
vanishes. 
Setting $\partial \omega/\partial k=0$
in Equation  (\ref{eq:om})---with $\omega_g$ assumed independent of $k$---
yields the following expression for $|K|$ at the turning point,
\begin{eqnarray}
K_{\rm QB}&=& g \quad {\rm (for\ WKB1.5)} \ .
\end{eqnarray}
Therefore, at a QB waves transition between
self-gravity- and pressure-dominated (Equations \eqref{eq:wkb15_1a} and \eqref{eq:wkb15_1b}).  
The ``QB path'' in the DRM is shown as an orange dashed line. 
It is identical to the $g(r)$ profile shown in Figure \ref{fig:basicstate} after normalizing to achieve $g_{\rm max}=2$.
\item
 {\it Lindblad resonance (LR)}: At a LR, the group velocity changes sign at  finite magnitude, producing 
  a kink in the DRM contour. This occurs at 
\begin{equation}
K_{\rm LR}=0 \ .
\end{equation}
At a LR waves switch between leading and trailing.
\end{enumerate}

\begin{figure}
	\centering
	\includegraphics[trim={0.9cm 0.5cm 0 0},clip,width=0.45\textwidth]{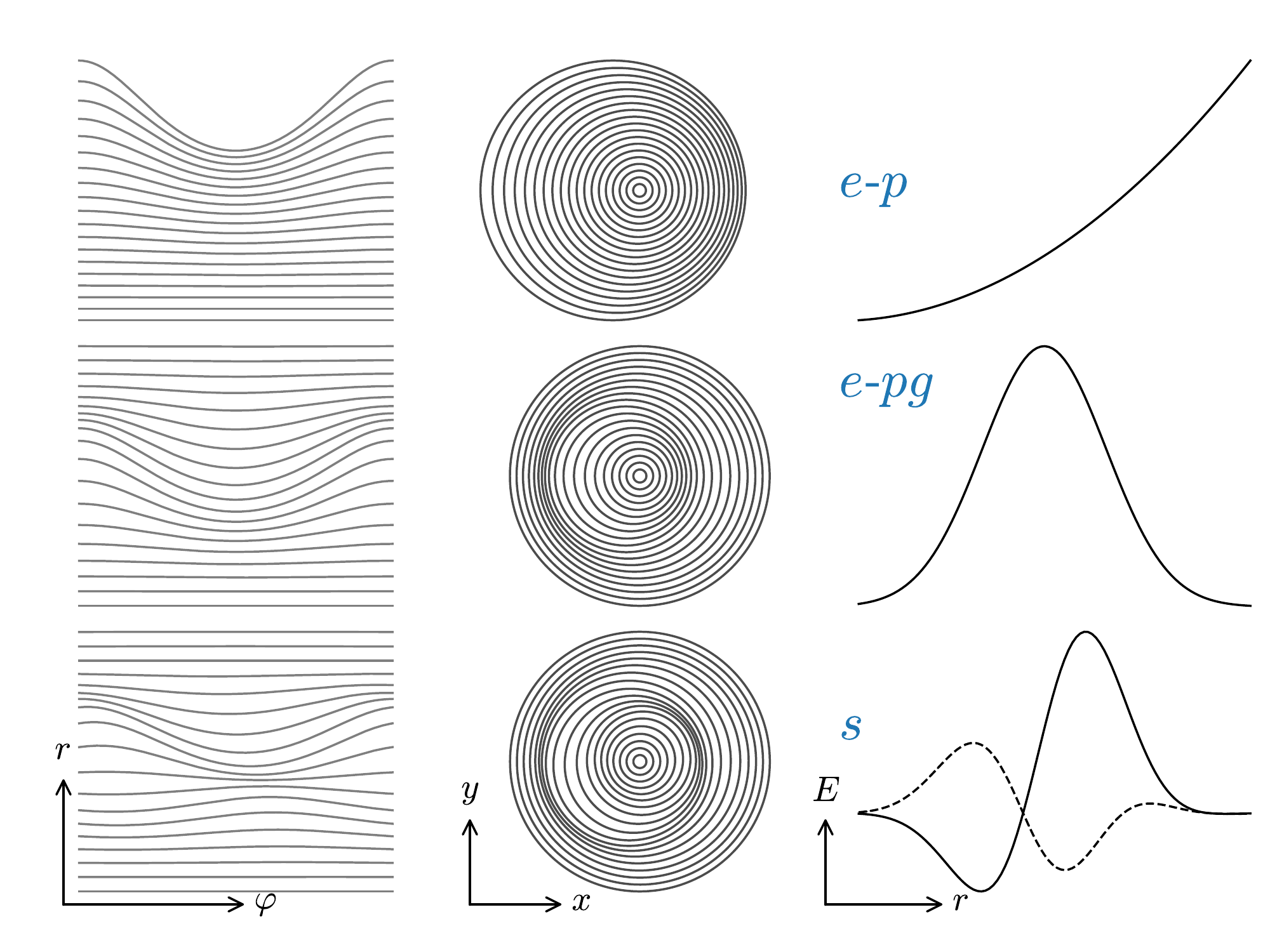}
	\caption{Illustration of the eccentric modes found in this work. From top to bottom rows, we show the elliptical-pressure ($e$-$p$), elliptical-pressure-gravity ($e$-$pg$), and spiral ($s$) modes.
	(\emph{Left col.}) The particle streamlines of the modes in polar coordinates. (\emph{Middle col.}) The streamlines in Cartesian coordinates. (\emph{Right col.}) Typical radial profiles of (real) eccentricity $E(r)$ are shown. The spiral mode has complex $E$, where the real and imaginary parts are shown as solid and dashed lines, respectively.}
	\label{fig:streamline_May2}
\end{figure}

The DRM in Figure \ref{fig:illustration} maps out a twin-peaked mountain, with peaks at $r\sim 0.05$ and $K\sim \pm 1$. We categorize the different kinds of modes as follows:
\begin{itemize}
	\item  Elliptical modes, or $e$ modes for brevity. These encircle both peaks of the mountain, and are symmetric in $K$. Therefore their normal modes are standing waves whose spatial appearance is of  apse-aligned elliptical rings (Figure \ref{fig:streamline_May2}). Two kinds of $e$ modes can be further distinguished, depending on whether their contours cross the QB paths at $|K|\lesssim 1$ or at $|K|\gtrsim 1$:
	
	\begin{itemize}
	\item Elliptical-pressure, or $e$-$p$ modes. These cross QB paths only at $|K|\lesssim 1$. In other words, they never cross QB paths in the portion of the DRM that corresponds to physical waves. Hence they are always on the short (pressure) branch, and bounce back and forth between two LRs. 
	\item Elliptical-pressure-gravity, or $e$-$pg$ modes. These cross the QB paths at least twice at $|K|\gtrsim 1$. In addition to the two LR reflections, they reflect at QBs, switching each
	time from short (pressure) to long (gravity) branches and back again.  	
	Although the distinction between $e$-$p$ and $e$-$pg$ modes might appear minor, there are at least two repercussions to crossing QB paths at $|K|\gtrsim 1$: the eccentricity $|E(r)|$ decays at large $r$ rather than rising exponentially (\S \ref{sec:outerdisk}), and the frequency is positive, meaning the modes are prograde (\S \ref{sec:frequecydiagramwkb15}).
	\end{itemize}
  
   \item Spiral, or $s$ modes. These encircle one of the two peaks without $K$ changing sign. A spiral mode that encircles the $K>0$ peak remains a trailing mode throughout its trajectory, and therefore its spatial appearance is of a one-armed trailing spiral (Figure \ref{fig:streamline_May2}). Similarly, the other spiral mode is a one-armed leading spiral. A spiral mode bounces back and forth between two QBs. In a numerical solution of the real equation of motion (Equation \ref{eq:basic3_2}), an $s$-mode pair appears as a doubly-degenerate eigenvalue, as in the bottom-right panel of Figure \ref{fig:eigensolutions}, with real eigenfunctions $E_1(r)$ and $E_2(r)$. But  one may equally well consider $E_1\pm iE_2$ to be the two eigenfunctions, one of which is purely trailing and the other purely leading.  The distinction between these two forms is only important when effects beyond the scope of this paper are considered, such as excitation or damping.
\end{itemize}
Our naming scheme differs from, e.g., \citetalias{2001AJ....121.1776T}, whose $p$ modes are essentially the same as our $s$ modes. We make this change for two reasons: (i) gravity and pressure are of comparable importance in $s$ modes (as well as in $e$-$pg$ modes), and therefore calling them $p$ modes might be misleading; and (ii) we distinguish between $s$ and $e$ modes because they can have different spatial appearance (Figure \ref{fig:streamline_May2}).

\subsection{Frequency Level Diagram  (WKB1.5)}
\label{sec:frequecydiagramwkb15}

The right panel of Figure \ref{fig:illustration} is the  ``frequency level diagram,'' an edge-on view
of the DRM in which solid curves depict $\omega$ along the QB- and LR-paths, and
the horizontal lines are the three modes' frequencies.
In this view, turning points are at the intersections.
Frequency level diagrams are analogous to energy level diagrams for a quantum mechanical particle in a potential. In the present case there can be two potentials at a given $r$, because $|K|$ has two separate  roots. Similar diagrams have been studied in, e.g., \citet*{1969ApJ...155..721L}.

The  $\omega_{\rm LR}$ and $\omega_{\rm QB}$ profiles play an important role
in determining the structure of the DRM and the frequencies of  modes.  For example, for the profiles shown in the figure, $s$ modes have frequencies between the peak of $\omega_{\rm LR}$ and the peak of
$\omega_{\rm QB}$, which implies they are prograde ($\omega>0$).
And their eigenfunctions are concentrated in the vicinity of those peaks, at $r\sim 0.03$. 
The $e$-$pg$ modes are also prograde, while the $e$-$p$ modes are retrograde and extend over a much
broader range of $r$. To see how the ``potentials'' in the frequency level diagram depend upon model parameters, we write \begin{eqnarray}
 \omega_{\rm LR}&=&\omega_p+\omega_g\vert_{k=0} \label{eq:omlr} \\
 &=& h^2\Omega\left(-0.6 + 1.0 g \right) \ ,  \label{eq:omlr2}
 \end{eqnarray}
 where the second equality is for our fiducial disk at $r\lesssim 1$ (Equation \eqref{eq:background_density}).
 Similarly,
 \begin{eqnarray}
  \omega_{\rm QB}&=&\omega_{\rm LR}+{1\over 2}{\mu^2\over h^2}\Omega \ \  {\rm (in\ WKB1.5)} 
  \label{eq:qb} \\
  &=& h^2\Omega\left( - 0.6 + 1.0 g+ 0.5g^2\right) \ .
  \label{eq:omqb2}
 \end{eqnarray}
One sees from these expressions that (i) $\omega_p$ is negative (as in Figure \ref{fig:illustration}); (ii) the peak
in $\omega_{\rm LR}$ is due to the peak in $g$ and scales in proportion to $g$ (neglecting the shift in peak location); and (iii) the peak 
in $\omega_{\rm QB}$ scales in proportion to $g^2$.

 \subsection{Quantum Condition}   
   
For a closed contour in the DRM to represent a standing mode, the phase accumulated over a complete
cycle must be an integer multiple of $2\pi$ \citepalias[e.g.,][]{1990ApJ...358..495S}.
There are two components to the accumulated phase: (i) 
 $\oint k(r,\omega)\,dr$, where the integration is over a cycle at fixed $\omega$; 
 and (ii) the sum of all of the phase changes at  turning points, including both QBs and
 LRs.
In Appendix \ref{sec:noteonlrqb} we derive an expression for (ii). 
Our derivation mostly follows \citetalias{1990ApJ...358..495S},  with one
notable exception:  
the phase change at slow-mode LRs, which has been treated incorrectly in the literature.
Our final result is that the turning points  contribute a total phase of $\pi$ for
\emph{all} slow modes, whether $e$-$p$, $e$-$pg$, or  $s$ mode.
 Therefore
the quantum condition reads
\begin{equation}
\label{eq:quantumcondition0}
\varointclockwise kdr = (2n+1)\pi = \pi,\,3\pi,\,5\pi, \cdots 
\end{equation}
with a clockwise integration to keep the integral positive (opposite to the direction of wave propagation in the DRM). In other words, standing modes are given by the contours in the DRM that enclose
area $(2n+1)\pi$ in the $k-r$ plane. Note that our convention is to plot DRM's with linear-log axes for $K$ vs. $r$, 
thus ensuring that  the area depicted  is $\oint (kr)d\log_{10} r$, and hence is quantized in proportion
to $(2n+1)\pi$.
We refer the integer $n$ as the quantum number of the mode. 

\section{WKB2 Analysis for Fiducial Suite}
\label{sec:wkb2analysis}

\begin{figure*}[htb!]
	\centering
	\includegraphics[width=\textwidth]{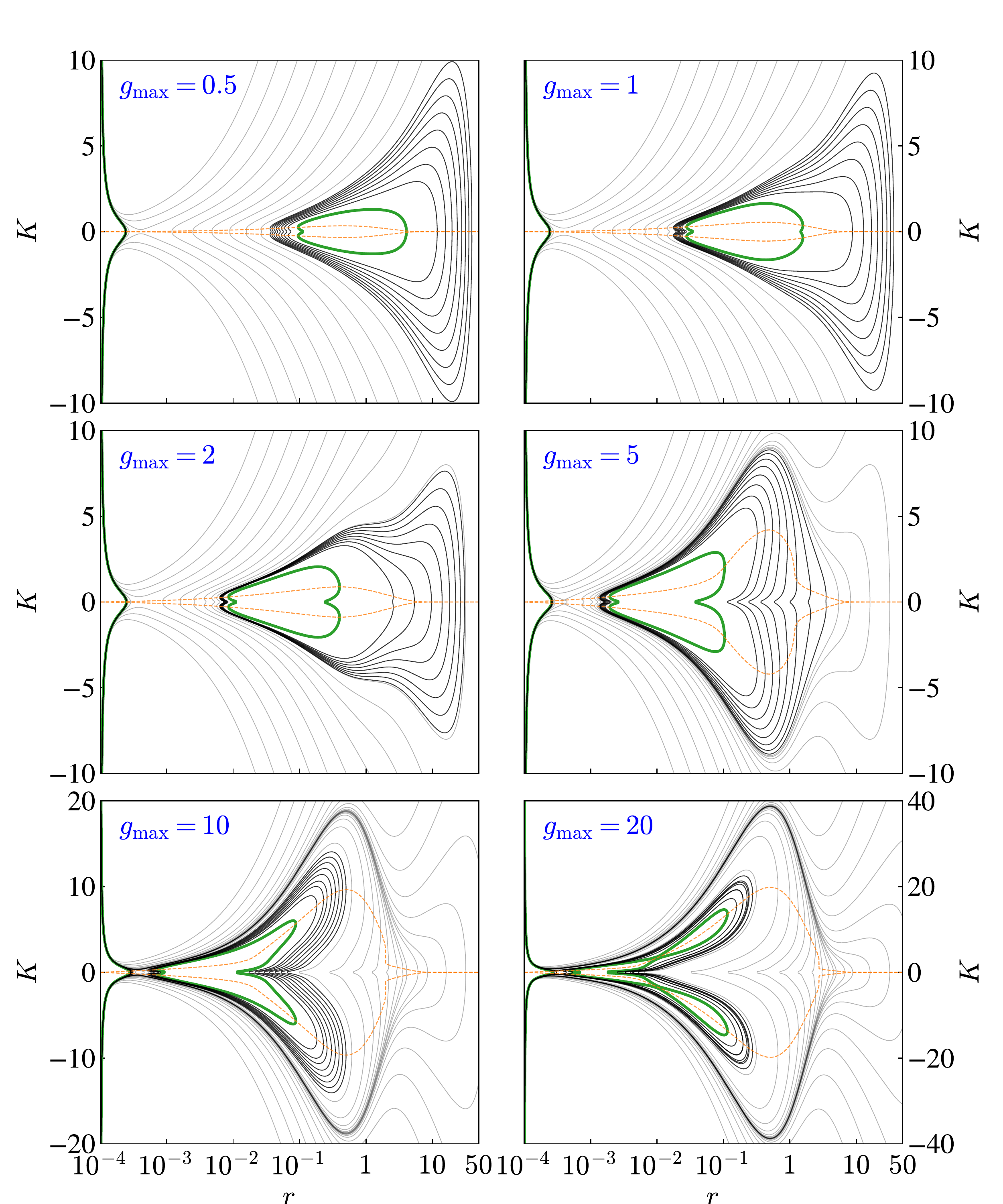}
	\caption{The WKB2 dispersion relation map based on Equation \eqref{eq:asymp1_1a} for disks with $g_{\rm max}=0.5$, 1, 2, 5, 10, and 20. The green contour is the one whose $\omega$ matches the highest frequency found numerically. The next 8 highest frequencies are shown in thick black, and the lighter contours beyond that are chosen arbitrarily to illustrate the structure of the DRM. The orange dashed curves are the ``QB paths" ($|K|=K_{\rm QB}$).}
	\label{fig:wkb_collection}
\end{figure*}

\subsection{Qualitative Comparison with Numerical Results}
\label{sec:wkb2qualitative}

\begin{figure}[htb!]
	\centering
	\includegraphics[trim={1.2cm 0cm 0 0},clip,width=0.47\textwidth]
	{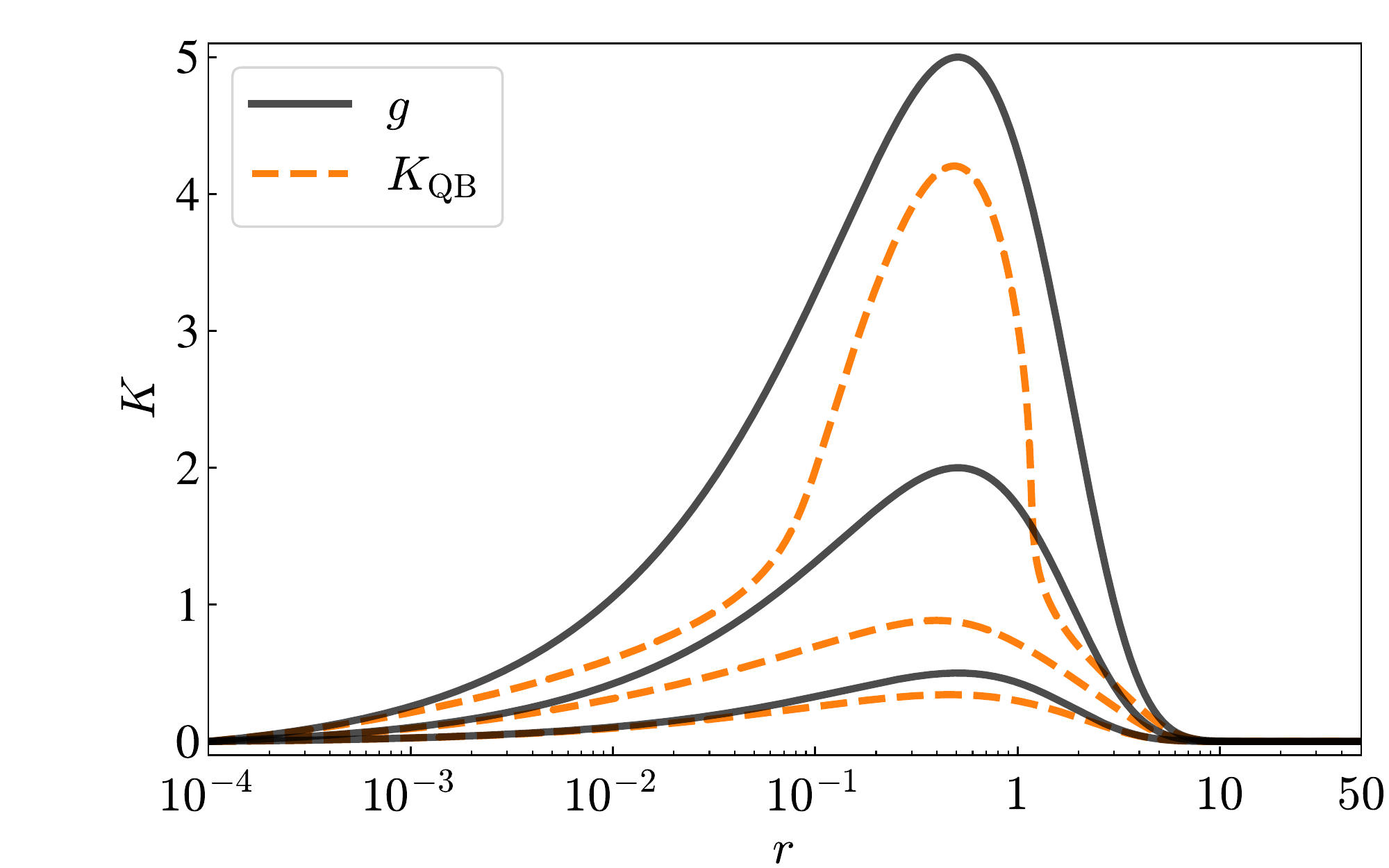}
	\caption{Comparison between the QB paths in WKB1.5 (black solid) and WKB2 (orange dashed). The top, middle, and bottom curves of each color corresponds to $g_{\rm max}=5$, 2, and 1, respectively. In WKB1.5, the QB path is $K=g$. In WKB2, we solve numerically for where the group velocity vanishes. This figure shows that the QB paths are similar---within a factor of $\sim 2$---in the two WKB approximations.}	\label{fig:kckqb}
\end{figure}

With an understanding of WKB1.5 in hand, we proceed to analyze the fiducial suite with the full WKB2 dispersion relation. Figure \ref{fig:wkb_collection} shows the six models' DRMs, 
as determined by WKB2 (Equations \eqref{eq:asymp1_1a}--\eqref{eq:omg}).  
We focus our discussion here on the fundamental mode, which is the mode with smallest number of nodes.
In a DRM, it is the contour with the smallest enclosed area that satisfies the quantum condition (green contour in figure; see also the discussion in \S \ref{sec:result_frequencies} below). At small $g_{\rm max}$, the fundamental mode is  an $e$-$p$ mode: it bounces between two LRs. At intermediate values, i.e., $g_{\rm max}=2$--$10$, it is an $e$-$pg$ mode, because
the QBs have split sufficiently to reflect the wave.   By $g_{\rm max}=20$, the contour has been
split into two, and now consists of two $s$ modes.
This progression with increasing $g_{\rm max}$ is simple to understand.
The value of $K$ along the QB path is  $K_{\rm QB}\approx g$; or, to be more precise, in WKB1.5 its
value is $K_{\rm QB}=g$, while in WKB2, $K_{\rm QB}$ differs slightly from $g$ (Figure \ref{fig:kckqb}).
Therefore increasing $g_{\rm max}$ simply pushes the two QB paths to higher $|K|$, thereby separating the mountain peaks in the DRM.
More physically, increasing $g_{\rm max}$ allows gravity-supported modes to propagate at 
increasingly short wavelengths, i.e., at larger $|K|$. 

\begin{figure*}[htb!]
	\centering
	\includegraphics[trim={0.7cm 0.5cm 0 0},clip,width=\textwidth]{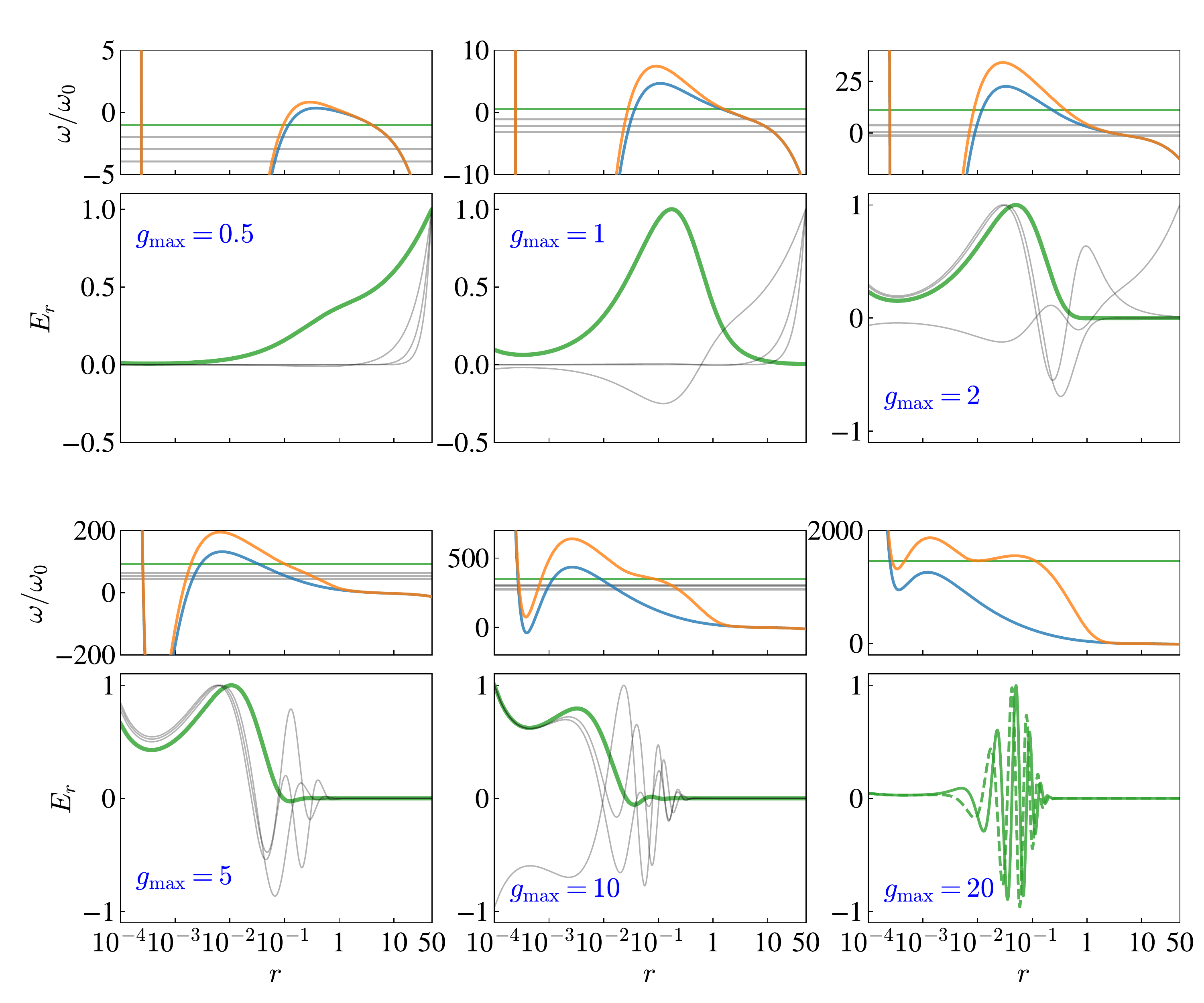}
	\caption{Same as Figure \ref{fig:eigensolutions}, which shows the eigenfrequencies and eigenfunctions for the six fiducial models. In addition, we show the frequency functions $\omega_{\rm LR}$ (blue) and $\omega_{\rm QB}$ (orange) on the top panel of each model.
	}
	\label{fig:Collection}
\end{figure*}

We turn now to the frequency level diagrams (Figure \ref{fig:Collection}). The  blue  and orange curves in the top panels are the frequencies along the LR and QB paths in the DRM.
The former, $\omega_{\rm LR}$, is determined by Equation (\ref{eq:omlr}); it is the same
as for WKB1.5 because $K_{\rm LR}=0$ in both cases. The latter, $\omega_{\rm QB}$, is determined by numerically solving for the turning points in WKB2;
although the result differs  from the WKB1.5 expression (Equation \eqref{eq:qb}), the difference is not large.
Also shown in Figure \ref{fig:Collection}
are the numerical eigenvalues (lines in the frequency level diagrams) and eigenfunctions, which are repeated here from Figure \ref{fig:eigensolutions}. One observes that, with the exception of the $g_{\rm max}=0.5$ case (to be discussed below), the
eigenfunctions are localized to the cavities predicted by  $\omega_{\rm QB}$ in the frequency level diagram. 
This figure shows that WKB2 provides a good description for the numerical solutions.

We may now also understand  the other three main features of the numerical solutions listed in \S \ref{sec:nr}:
(i) Increasing $g_{\rm max}$ causes the fundamental mode to have shorter wavelength because it 
pushes the DRM mountain peaks to higher $|K|$.
(ii) The fundamental mode is  the one that most closely encircles the mountain peak (or peaks)
while enclosing the smallest area allowed by the quantum condition. 
Higher harmonics enclose more area and hence are  further down the mountain. As a result, the fundamental mode always has the highest $\omega$, and higher harmonics have lower values.  (iii) At $g_{\rm max}=0.5$, both $\omega_{\rm LR}$ and $\omega_{\rm QB}$ are negative and close to each other. This may be seen in Figure \ref{fig:Collection}, and explicitly 
from Equations (\ref{eq:omlr2}) and (\ref{eq:omqb2}), with the caveat that
the latter equation only approximately holds in WKB2.  Therefore, the fundamental mode is retrograde.
Furthermore,  Equations (\ref{eq:omlr2}) and (\ref{eq:omqb2}) also show that increasing $g_{\rm max}$ pushes the peaks of
$\omega_{\rm LR}$ and $\omega_{\rm QB}$ to higher values. 
Therefore, increasing $g_{\rm max}$ pushes the fundamental mode's frequency to increasingly 
positive values. 

\subsection{Quantitative Comparison of Mode Frequencies}
\label{sec:result_frequencies}

In this section we compare the numerically computed eigenfrequencies to 
the WKB2 predictions. Before discussing the results, we first describe our method to compute the frequency from WKB2. For a given frequency, we obtain the coordinates along its contour in the DRM and calculate the enclosed area using Green's theorem. We then find each contour whose area is an odd integral multiple of $\pi$ as in Equation \eqref{eq:quantumcondition0}. The advantage of using an area method over performing the phase integration $\oint k dr$ \citepalias[as done by][]{1990ApJ...358..495S} is that we do not have to compute turning point locations explicitly and then invert $\omega(k)$ into $k(\omega)$ for a given $\omega$. Therefore, the area method is  more convenient for non-trivial dispersion relations, such as WKB2.

\begin{figure}[t]
	\centering
	\includegraphics[trim={0.1cm 0.2cm 0cm 0.2cm},clip,width=0.5\textwidth]
	{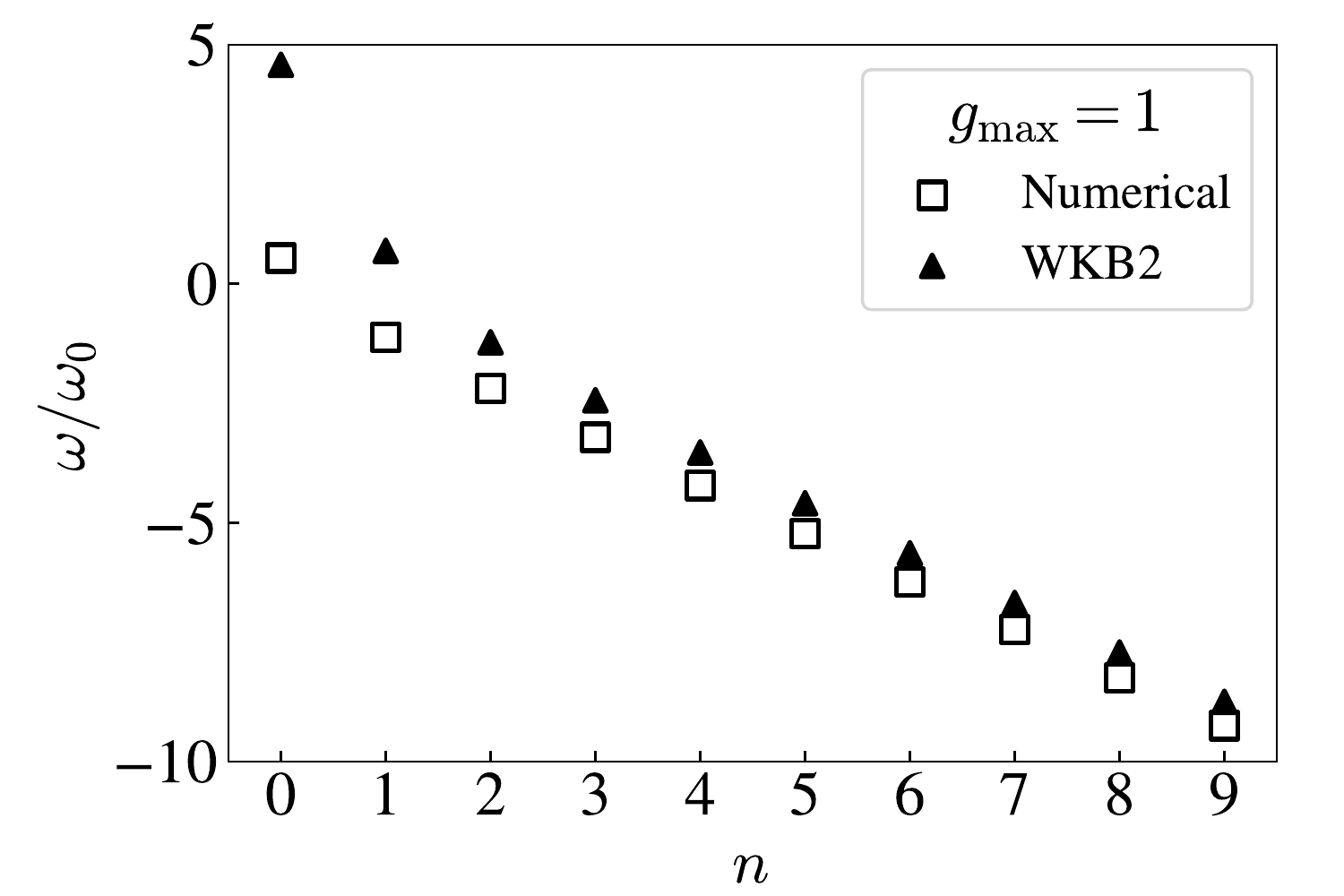}
	\caption{Comparison of eigenfrequencies: numerical (open squares) and WKB2 predictions (solid triangles). We set $g_{\rm max}=1$. The horizontal axis is the quantum number $n$ that is used for the WKB predictions (Equation \ref{eq:quantumcondition0}). The fundamental mode is an $e$-$pg$ mode and the rest are $e$-$p$ modes.}
	\label{fig:quantumconditionloopg1}
\end{figure}

\begin{figure}[t]
	\centering
	\includegraphics[trim={0.1cm 0 0cm 0.2cm},clip,width=0.5\textwidth]
	{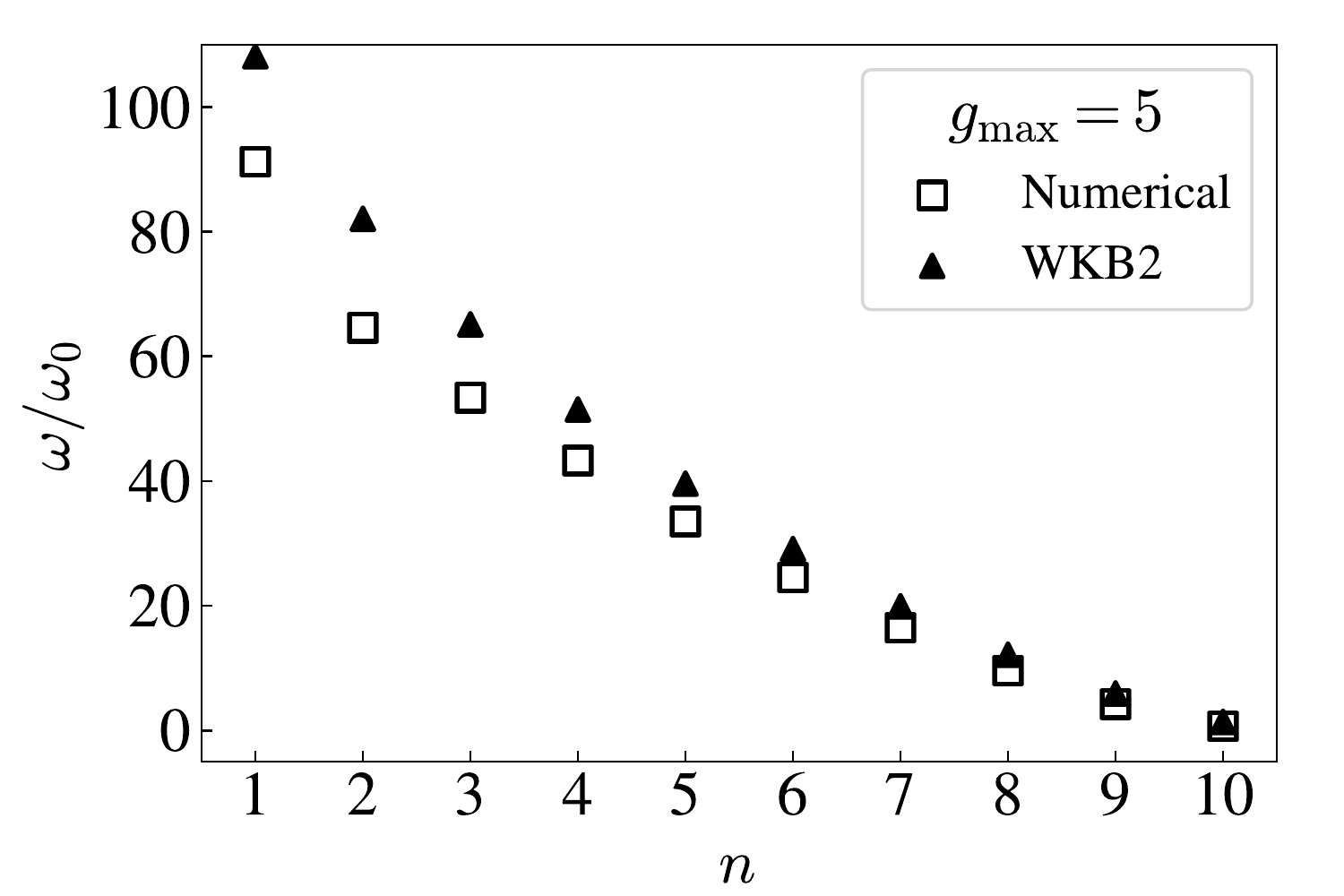}
	\caption{Same as Figure \ref{fig:quantumconditionloopg1}, but for a $g_{\rm max}=5$ disk. The first 10 modes are shown, which are all $e$-$pg$ modes.}
	\label{fig:quantumconditionloopg5}
\end{figure}

\begin{figure}[t]
	\centering
	\includegraphics[trim={0.1cm 0cm 0cm 0.2cm},clip,width=0.5\textwidth]{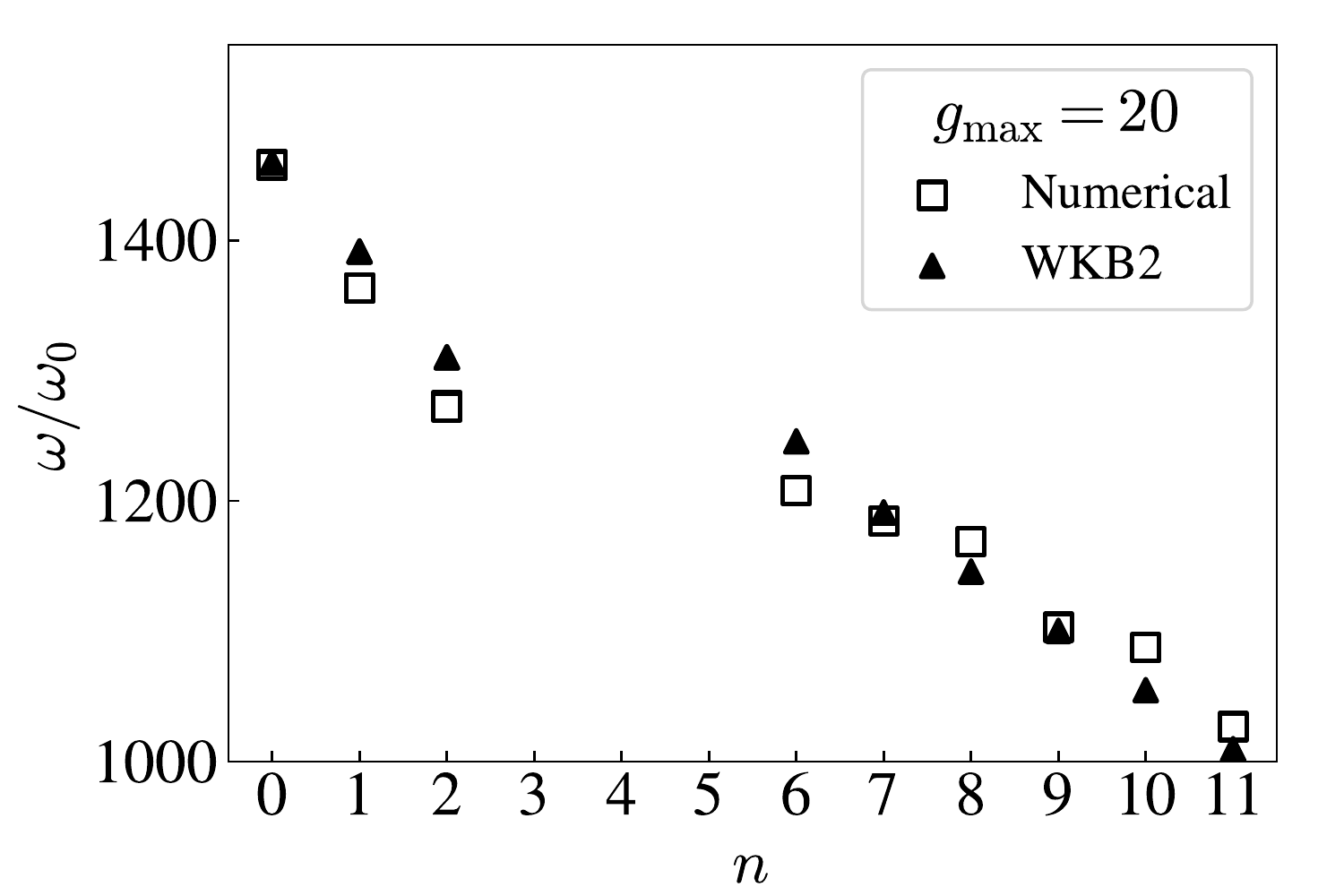}
	\caption{Same as Figure \ref{fig:quantumconditionloopg1}, but for $g_{\rm max}=20$. The first 3 pairs ($n=0-2$) are $s$ modes. The $n=6-11$ modes are $e$-$pg$ modes. }
	\label{fig:quantumconditionloopg20}
\end{figure}

We show the results in Figures \ref{fig:quantumconditionloopg1}, \ref{fig:quantumconditionloopg5}, and \ref{fig:quantumconditionloopg20} for disks with $g_{\rm max}=1$, 5, and 20, respectively. Our results show very good agreement between numerical values and WKB2 estimates of the eigenfrequencies. As seen in the figures, the error decreases as the number of nodes increases, i.e., towards large $n$. On the other hand, the modes with large wavelength ($K \sim 1$) have larger errors because of the inaccuracy of the dispersion relation in this regime.

For $g_{\rm max}=5$ (Figure \ref{fig:quantumconditionloopg5}), the first 10 modes are shown, which are all $e$-$pg$ modes. Note that in this case we assign the highest numerical frequency to $n=1$ rather than $n=0$, and correspondingly shift the other numerical frequencies from  $n$ to $n+1$. Although this assignment might seem arbitrary, we do it because there is no contour in the WKB2 DRM with area $\pi$. That is because for $g_{\rm max}=5$, contours with area smaller than that of the fundamental mode transition from enclosing both mountain peaks to enclosing the two peaks separately.  At the transition point, the enclosed area changes discontinuously from larger than $\pi$ to smaller than $\pi$. Further support for this assignment of $n$ values is provided by  the eigenfunction of the fundamental mode in $g_{\rm max}=5$ (see Figure \ref{fig:Collection}), which has one node.

For $g_{\rm max}=20$ (Figure \ref{fig:quantumconditionloopg20}), the first 3 data points are the degenerate trailing/leading pairs of $s$ modes. Each pair of $s$ modes shares the same frequency and hence the same $n$. The next mode is an $e$-$pg$ mode with $n=7$. The gap in $n$ (without any numerical or WKB modes) appears when the quantized contours in the DRM transition from enclosing one mountain peak to enclosing both of them, as described above for the $g_{\rm max}=5$ case.

\subsection{Eccentricity in the Outer Disk}
\label{sec:outerdisk}

\begin{figure}
	\centering
	\includegraphics[trim={0.1cm 0.2cm 0cm 0.2cm},clip,width=0.47\textwidth]{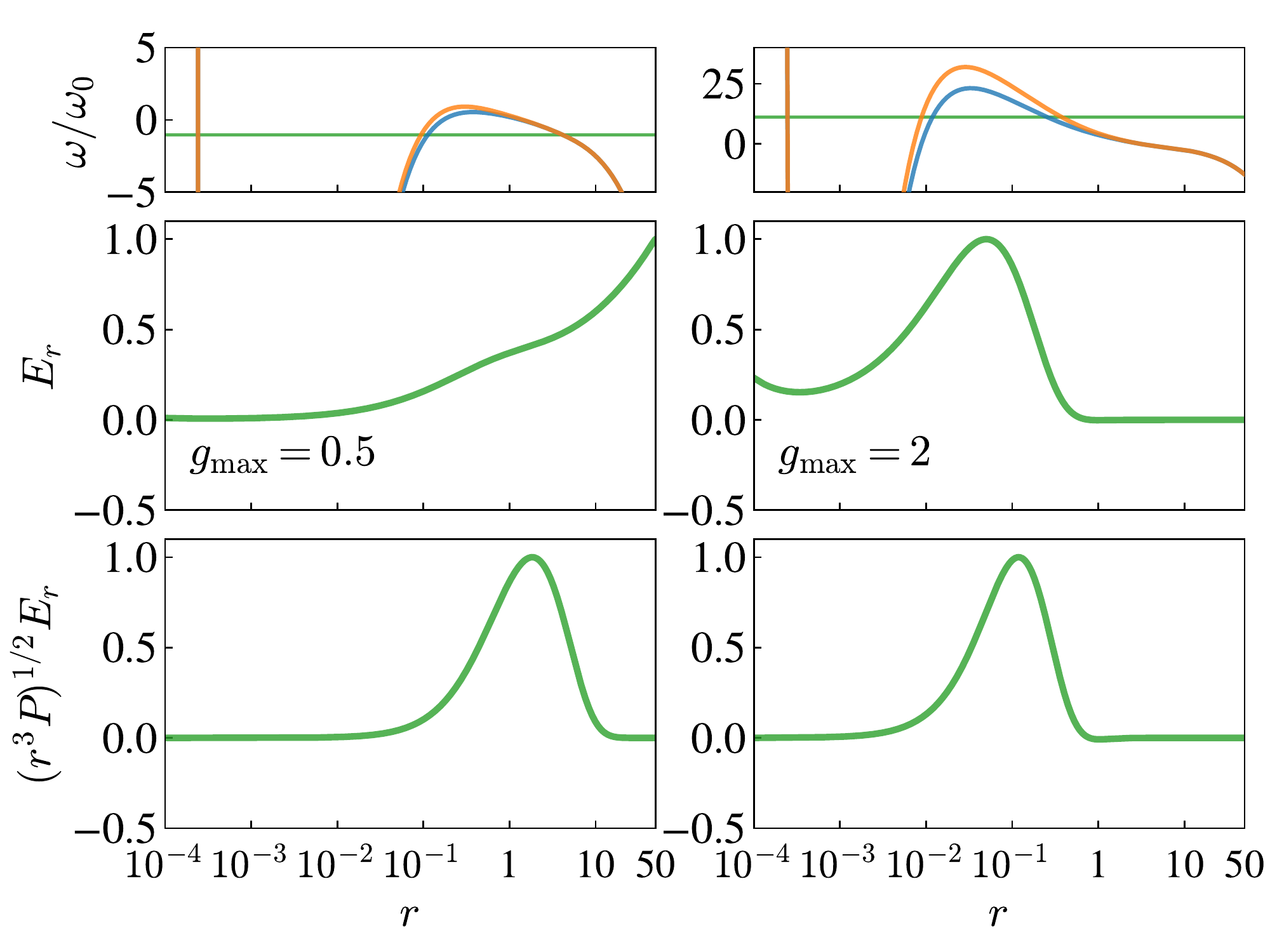}
	\caption{The fundamental modes for the $g_{\rm max}=0.5$ and 2 cases. The top two panels
		for each case are replotted from Figure \ref{fig:Collection}. The bottom panel shows the eigenfunction in 
		terms of the normal form variable, demonstrating that it is confined by the cavity predicted
		by the frequency level diagrams in the top panels.}
	\label{fig:eig2}
\end{figure}

In this subsection, we address why the eigenfunctions in the $g_{\rm max}=0.5$ case are seemingly
unconfined by their wave cavities as predicted by the frequency level diagram (Figure \ref{fig:Collection}). Rather, the eigenfunctions continue to rise towards the outer disk, far beyond the exponential
cutoff of the disk at $r\sim 1$.  A similar effect can be seen in the $g_{\rm max}=1$ case for the $n>0$  modes. 
The reason for this behavior is that in deriving WKB2, one finds that
$(r^3P)^{1/2}E\propto \exp i\int kdr$. In other words, the relevant (``normal form'') dynamical variable is not $E$, but
$(r^3P)^{1/2}E$ (see discussion below Equation \eqref{eq:omg} and  Appendix \ref{sec:derivationofDR2}).  
The left panels of Figure \ref{fig:eig2}  replot the $g_{\rm max}=0.5$ fundamental mode, this time
showing also in the bottom panel the normal form eigenfunction. It is clear that the normal form
eigenfunction is confined by the predicted wave cavity.  An analogous effect can be seen at small
$r$ (right panels of Figure \ref{fig:eig2}).

Since the eccentricity $E$ is  potentially observable in a real disk, let us
examine the condition under which is rises outwards.
From $E\propto (r^3P)^{-1/2}\exp i\int kdr$,  we see that $|E|$ rises outwards when the exponential rise in the prefactor---caused by the exponential drop in $\Sigma$ and hence in $P$ at $r\gtrsim 1$---dominates the
decay caused by $k$. 
To be more precise, we focus on the behavior of the eigenfunction in the vicinity of the outer turning point $r_{\rm tp}$. 
For example, in the left panel of Figure \ref{fig:eig2}, $r_{\rm tp}\sim 5$, where the horizontal green line intersects the orange $\omega_{\rm QB}$ curve. 
We employ WKB1.5,
\begin{align}
\label{eq:wkb15}
\omega = \omega_p + \omega_g - \frac{1}{2}K^2 h^2 \Omega + \mu |K| \Omega \ ,
\end{align}
which is adequate for this purpose, and define $\Delta\omega\equiv \omega-\omega_{\rm QB}$, in which case
\begin{equation}
\Delta\omega = -{h^2\Omega\over 2}\left( K-{\mu\over h^2}\right)^2 \ ,
\end{equation}
where for simplicity we drop the absolute value on $K$, which is equivalent to focusing on the 
trailing branch.

At the turning point, $\Delta\omega=0$ because $K=K_{\rm QB}=\mu/h^2$ there.
At $r>r_{\rm tp}$, $\Delta\omega > 0$ and hence $K$ is complex:
\begin{equation}
K={\mu\over h^2}\pm i\sqrt{\frac{2\Delta \omega}{h^2\Omega}} \ .
\end{equation}
Therefore, the amplitude of $|E|$ at $r>r_{\rm tp}$ is primarily governed by
\begin{equation}
|E|\propto P^{-1/2}e^{-\int dr \sqrt{\frac{2\Delta \omega}{h^2\Omega}}} \ , \  \ {\rm at\ } r>r_{tp}
\end{equation}
To proceed further, we take $P\propto e^{-r}$  (as in Equation \eqref{eq:background_density} for $p=1$ after
dropping power-law terms), and we evaluate $\Delta\omega$ by Taylor expanding at 
$r_{\rm tp}$:  $\Delta\omega=-(r-r_{\rm tp}){d\omega_{\rm QB}\over dr}\vert_{\rm tp}$, in which case
\begin{eqnarray}
|E|&\propto& \exp\left[{-{2\over 3}\beta  (r-r_{\rm tp})^{3/2}+{r-r_{\rm tp}\over 2}}\right] \ ,
\end{eqnarray}
where
\begin{eqnarray}
\beta &\equiv& \left(-{2\over h^2\Omega}{d\omega_{\rm QB}\over dr} \right)^{1/2}\Bigg|_{\rm tp} \ . 
\end{eqnarray}
When $\beta \gtrsim 1$, the first term in the exponent dominates, and $|E|$ decays outwards. And 
when $\beta\lesssim 1$, the second (pressure) term dominates and $|E|$ rises exponentially. 
Finally, we may relate $\beta$ to $g$ in an approximate way by considering the last
term in Equation  \eqref{eq:omqb2}.  We see
that exponential
rise occurs when $g\vert_{r_{\rm tp}}\lesssim 1$; and otherwise,  decay occurs, as is consistent with the eigenfunctions in Figure \ref{fig:Collection}. 

\section{Dependence on the Disk Profile}
\label{sec:diskprofiles}

\begin{figure}[t!]
	\centering
	\includegraphics[trim={0.8cm 0cm 0 0},clip,width=0.47\textwidth]{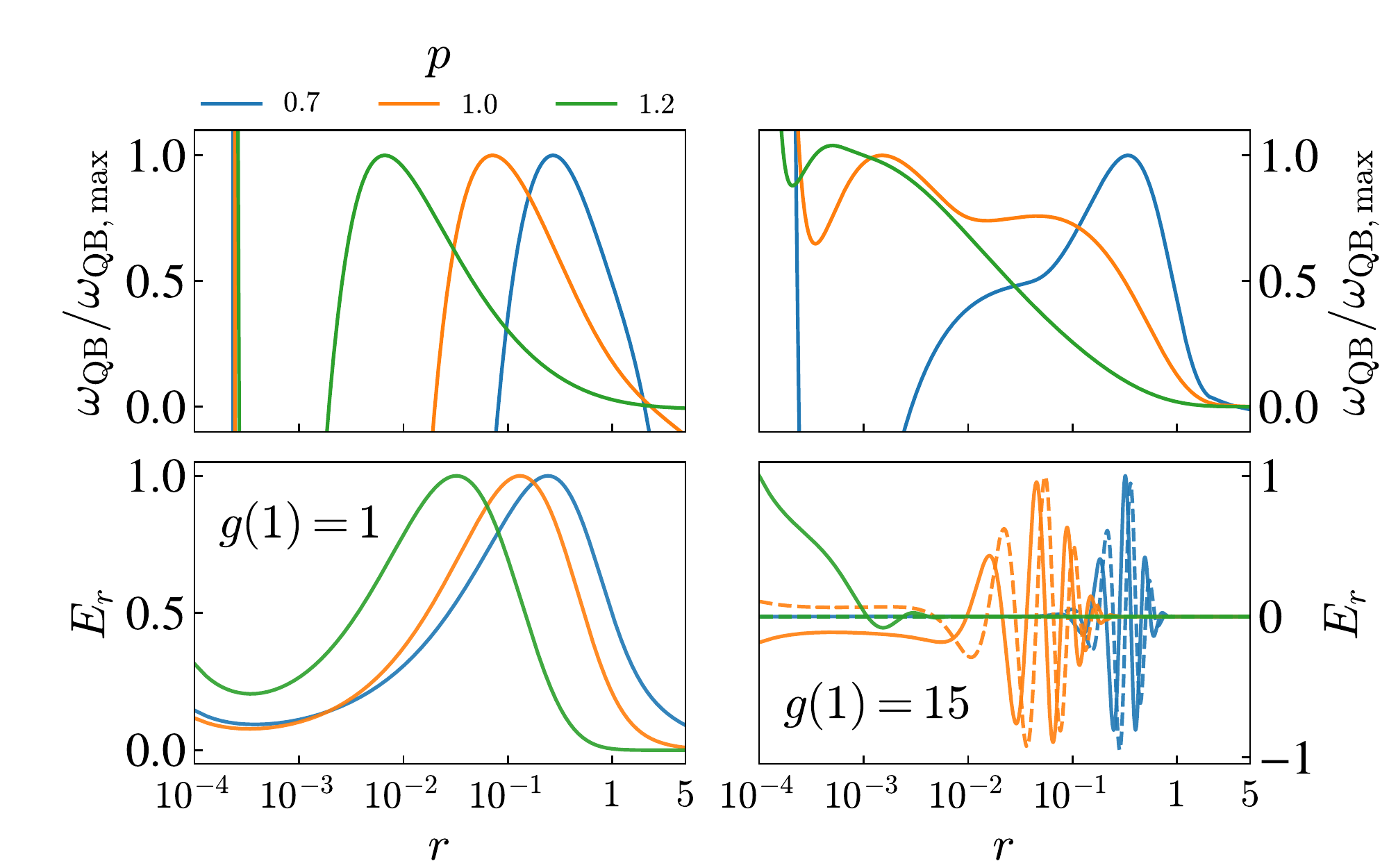}
	\caption{Comparison of disks with different density slope parameter $p$, and different
			normalizations of the $g$ profile. The left and right columns show the results for $g(r=1)=1$ and $15$, respectively. The blue, orange, and green curves correspond to the values of $p=0.7$, 1.0, and 1.2, respectively. The top panels show the $\omega_{\rm QB}$ curves, normalized to the maxima. The bottom panels show the eccentricity profile (real part of $E$) of the fundamental, i.e., highest-frequency, modes. The dashed curves on the bottom-right panel correspond to the second fundamental mode at the same frequency ($s$ modes). For clarity, the outer disk ($r>5$) is not shown.}
	\label{fig:diskcompareslope}
\end{figure}

We move on to examine the dependence of eccentric modes on the disk profiles in this section. Figure \ref{fig:diskcompareslope} shows the fundamental modes for disks with different density-slope parameter $p$ of the surface density (Equation \eqref{eq:background_density}), and different normalizations of the $g$ profiles. A shallower density slope (and hence a sharper edge) moves the modes outward because the peak of $\omega_{\rm QB}$ (upper panels) shifts in the same way (see \S \ref{sec:frequecydiagramwkb15}). 
We find similar results for both $g(1)=1$ and $g(1)=15$, which have $e$ and $s$ fundamental modes, respectively. For $g(1)=15$ and $p=1.2$, $s$ modes are not supported.

\section{Discussion}
\label{sec:discussion}

\subsection{Comparison With Previous Work}
\label{sec:previouswork}

\citetalias{2001AJ....121.1776T} found two types of modes in his pressureless disks with softened 
self-gravity: prograde $p$ modes and retrograde $g$ modes (in his naming convention). 
As described in \S  \ref{sec:mp}, his $p$ modes correspond to our $s$ modes, which are also prograde.  In his case these modes are trapped by QBs caused by gravitational softening, as shown in his DRM (his Figure 1). His $g$ modes only exist in narrow ring profiles or if an external source of precession is imposed, and hence we do not find them in the present work. Conversely, he  does not find our $e$ modes  because he does not include pressure forces explicitly. Although he does use gravitational softening as a crude model for pressure, it does not produce the  LR-crossing contours that appear in our DRM, as can  be seen by again comparing  his DRM with ours. 

Slow modes in self-gravity-only disks have also been examined via Laplace-Lagrange theory.  \citet*{1999ASPC..160..307S} studied the normal modes by modeling the disk as a set of
non-intersecting orbits. However, such a model does not reproduce the behavior of
gas or particle disks, as explained in \citetalias{2001AJ....121.1776T}. \citet{2003ApJ...595..531H}  showed that the results of Laplace-Lagrange theory, modified to account for the disk's finite thickness, agree with the fluid approach of \citetalias{2001AJ....121.1776T}.

\citet{Pap2002} was the first to formulate the slow mode linear eigenvalue problem  in the presence of both pressure and self-gravity. He  numerically solved for the modes in two disk models, and also explained based on  the Lin-Shu dispersion relation why more massive disks have higher frequency modes, consistent
with one of our findings.
   We note that \citet{Pap2002} does not find any $s$ modes as his $g$ parameter is not large enough. 
\cite{2016MNRAS.458.3221T} also explored numerically disks with pressure and self-gravity, and also found that increased mass leads to higher frequency modes.

\citet{2006MNRAS.368.1123G}, \citet{2008MNRAS.388.1372O}, and \citet{2009MNRAS.400.2090S} \citep[and see  also,][]{2016MNRAS.458.3221T} explored slow modes in disks with pressure but  no self-gravity, both theoretically and numerically. They showed that the eccentricity equation
can be transformed into a Schr\"{o}dinger-like equation, and therefore
its solutions can be understood and analyzed graphically.
This approach is equivalent to our frequency-level diagrams.
We may see this  explicitly from our pressure-only WKB2 (or WKB1.5) dispersion relation $\omega=-{c^2\over 2\Omega}k^2+\omega_p$ (Equation \eqref{eq:asymp1_1a}), which is the same as for the Schr\"{o}dinger equation with potential $\omega_p$---aside from the $r$-dependence of $c^2/\Omega$. One could even remove the $r$-dependence by transforming coordinates \citep[as done by][]{2008MNRAS.388.1372O}, although we did not find that step useful
 for this paper. Using different dynamical variables, \citet{2009MNRAS.400.2090S} obtained an analogous expression for $\omega_p$. In Paper II, we explore the pressure-only case in more detail, and show in particular that slow
modes are typically trapped in disks that have more realistic boundary conditions than those
used in the aforementioned papers.

A particularly intriguing result is that of \citet{2015MNRAS.448.3806L}, who simulated disks with self-gravity and pressure using a full hydrodynamics code. He found an instability when he made the equation of state locally isothermal
(rather than locally adiabatic as in the present paper), and when the the surface density had
a  ``bump'' within which $1<Q<2$. The result of the instability was a growing trailing $m=1$ spiral ($K>0$). As he showed, the locally isothermal equation of state leads to an instability because it does not conserve angular momentum.  He also used a leading-order WKB theory (i.e.,  WKB0) to correctly predict the growth rate of the unstable mode. In our nomenclature, the unstable mode is an $s$ mode. As we shall show elsewhere, a simple modification to our theory  predicts that such an instability should occur in more general disk profiles  as long as the conditions for $s$ modes are satisfied. 

\citet{2016MNRAS.458.3918Z}, in a 
numerical study of vortices in self-gravitating disks,
found a case in which the vortex dispersed in a strongly-self-gravitating disk (local $Q \sim 1$), and then caused the disk to become eccentric. A prograde $m=1$ mode was found to be precessing slowly compared to the local Keplerian speed of the initial vortex. We speculate that this was an $e$-$pg$ mode, although further work is needed to verify this. 

\subsection{Additional Comments}
\label{sec:discussion_connection}

\paragraph{Angular Momentum Flux of $s$ modes} The $s$ mode is a one-armed spiral. 
One might expect that a spiral would carry a non-zero flux of angular momentum.
But in fact an $s$ mode carries zero flux, because it is a standing
wave that is a superposition of  two branches (long and short) that carry equal and
opposite fluxes. 
The flux in a   branch is  $\mathcal{F} = \mathcal{J} v_g$ \citep[][]{1972MNRAS.157....1L,1978ApJ...222..850G}, where $v_g$ is the group velocity 
\citep{1969ApJ...158..899T,1970ApJ...160...99S} and $\mathcal{J}$ is wave angular momentum density (\citealt{1979ApJ...233..857G}; \citetalias{1990ApJ...358..495S}), which for slow $m=1$ modes
is
\begin{align}
\label{eq:toymodel1_5b}
\mathcal{J}  = -\pi r^3 \Sigma \Omega |E|^2 \ ,
\end{align}
which is equivalent to the density of angular momentum deficit \citep[e.g.,][]{2016MNRAS.458.3221T}. Since the group velocity takes opposite sign for short/long waves, the total flux vanishes as claimed.

\paragraph{Stability}
\label{sec:discussion_stability}

The $m=1$ slow modes studied in this paper are all stable.  But do $m=1$
modes remain stable if the modes are not strictly slow, e.g., if corotation
lies within the disk?  
A number of disk instabilities have been considered in the literature that operate when a disk 
is close to, but not quite, Toomre-unstable. 
First, disks can be unstable via over-reflection of a wave near corotation, 
leading to SWING and WASER instabilities  \citep{1976PhDT........26Z,1976ApJ...205..363M,1978ApJ...222..850G,1981seng.proc..111T} \citep[for an introduction see e.g.,][]{1992pavi.book.....S}.
 But this  process is only effective when the QB is close to corotation, and hence
  should not operate for $m=1$ modes when their corotation is far away---whether or not it lies
  within the disk. 
  And second, the  SLING instability 
 (\citealt{1989ApJ...347..959A}, \citetalias{1990ApJ...358..495S}) is driven by the indirect
 potential---which vanishes in the slow mode
 approximation, and hence might drive instability if one relaxed that approximation. 
  But thus far the SLING
 instability has only been studied in disks with very sharp outer cutoffs. We suspect that it will
  not operate in the more realistic case of an exponential cutoff to the disk. Nonetheless, 
 further work must be done to test this speculation.

\paragraph{``Very Long" Waves}
\label{sec:discussion_verylong}
For the Lin-Shu dispersion relation, as well as in WKB0 and 1.5, there are two roots
for $|k|$: the long and short branches. 
In WKB2, the $k$-dependence of $\omega_g$ (Equation \eqref{eq:omg}) leads to an additional
root at very small $|k|$, called the ``very long" branch \citep{1978ApJ...226..508L,1989ApJ...338..104B,2014dyga.book.....B}. But this root appears to have little significance for eccentric modes because it always occurs at $|kr|\lesssim 1$. Hence it has little effect on  the quantum condition, and also cannot support a mode because the WKB approximation becomes invalid at those small $k$'s.

\paragraph{Three-dimensional Effects}

We have  considered only 2D disks without any vertical structure. 
But in a 3D disk, a fluid parcel cannot maintain vertical hydrostatic equilibrium while on an eccentric orbit, 
which results in the eccentricity equation acquiring an extra term \citep{2008MNRAS.388.1372O,2016MNRAS.458.3221T,2018MNRAS.477.1744O}. 
Although the magnitude of this extra term is comparable to the other pressure terms, it
tends to switch the sign of $\omega_p$ within the outer cutoff radius, which can have 
significant consequences. 
In particular, if $\omega_p$ is positive (rather than negative, as in this paper), 
the $\omega_{\rm LR}$ curve in the frequency level diagram (such as Figure \ref{fig:illustration}) would continue to rise
to small $r$,  removing the innermost LR from the disk.  In other words, the $e$-modes could
no longer exist because they could not bounce back from an inner LR. 
Quantitatively, in the 2D case, $\omega_p$ is always negative when $0<p<2$ and $0<q<2$ for $\gamma=1.5$. Whereas from the 3D equations of Ogilvie
and collaborators, one requires $p+q < 2.3$, for $\gamma=5/3$. The former criterion
is satisfied by most reasonable protoplanetary disk models, while the latter may not be.

Although 3D effects could significantly alter  our results, there is also a caveat: the derivation of
the 3D eccentricity equation assumes  that viscosity is big enough 
to force $E$ to be independent of height. If the viscosity is small, the theory becomes
``very difficult to describe.'' \citep{2008MNRAS.388.1372O}.  
We leave such difficulties to future work. 

\subsection{Implication for Observations}

Eccentric disks can cause wide lopsided feature in disks \citep[e.g.,][]{2012ApJ...760..119H,2013A&A...553L...3A,2016MNRAS.458.3918Z,2016MNRAS.458.3927B}. By studying deprojected maps of dust continuum observations, one may measure the disk eccentricity by fitting an ellipse and comparing its focus location to the known stellar location, as was done by \citet{2018ApJ...860..124D}. Azimuthal brightness variation \citep{2016ApJ...832...81P} can also be expected due to the slower velocity near the apocenter of a fluid orbit.

We also propose that the eccentricity profile and its gradient can be a proxy for measuring the disk mass, through the gravity parameter $g$. More wavy eccentricity profiles are expected for more massive disks (Figure \ref{fig:eigensolutions}) and $|E|$ generally drops outside the exponential cutoff radius (\S \ref{sec:outerdisk}). We note that the linear theory does not give the distortion amplitude, which depends on the initial condition. 

At small amplitudes, the relative azimuthal velocity change is $E$. Gas kinematic signatures should be within current observational capability of ALMA, as has been demonstrated recently by \citet{2018ApJ...860L..13P} for possible planet-disk interaction.

\section{Summary}
\label{sec:conclusion}

\begin{table*}[t]
	\centering
	\begin{tabular}{lccc}
		\hline\hline
		mode & $e$-$p$ & $e$-$pg$ & $s$\\
		\hline
		branch & short & long+short & long+short\\
		LRs $(K=0)$? & yes  & yes & no\\
		$Q$-barrier? & no & yes & yes\\
		range of $g$ & any & $g \gtrsim 5$ & $g \gtrsim 10$\\
		frequency\footnote{Not all modes exist in the same disk.} & lowest & intermediate & highest \\
		studied by:\footnote{Non-exclusive list} & 
		\cite{2006MNRAS.368.1123G} &
		\citet{Pap2002} & \citetalias{2001AJ....121.1776T}; \citet{2015MNRAS.448.3806L} \\
		\hline
	\end{tabular}
	\caption{Summary of eccentric modes in a self-gravitating fluid disks}
	\label{tab:modes}
\end{table*}

\begin{enumerate}
	
	\item We solved numerically for slow eccentric ($m=1$) modes in disks with both self-gravity and pressure, focusing primarily on a suite of six model disks with varying ratio of self-gravity to pressure (parameterized by $g=\mu/h^2$). The eigenvalues and eigenfunctions for the suite are displayed in 
	Figure \ref{fig:eigensolutions}. We also developed a simple and efficient
	 numerical eigen-solver that improves upon what has appeared in the literature (Appendix \ref{sec:methodofsolution}).
	
	\item We derived the second order WKB theory [WKB2] that describes these modes
	even when their wavelengths are nearly as large as the radial distance to the star (Appendix \ref{sec:derivationofDR2}). And we derived their quantum condition, correcting errors that had appeared in the literature (Appendix \ref{sec:q})
	
	\item We showed  in the context of a slightly simplified theory [WKB1.5]
	that the properties of the modes are simply understood with the help of a dispersion relation
	map (DRM) and its corresponding frequency level diagram (Figure \ref{fig:illustration}). 
	
	\item We found three different kinds of modes, which we dub $e$-$p$, $e$-$pg$, and $s$ modes.
	Table \ref{tab:modes} summarizes their properties.

	\item We applied our full WKB2 theory to the modes found in the suite of six models, 
	finding agreement between theory and numerical solutions (Figure \ref{fig:Collection}). 
	We also explained, with the help of the DRM, how the fundamental mode (i.e., 
	largest wavelength/highest frequency mode) depends on $g_{\rm max}$:  at $g_{\rm max}\lesssim 1$, 
	it is a retrograde $e$-$p$ mode, at  $2\lesssim g_{\rm max}\lesssim 15$ it is a prograde $e$-$pg$ mode, and at $g_{\rm max}\gtrsim 20$ it becomes two degenerate $s$ modes.
	
	\item We  tested the theory quantitatively by computing the eigenvalues from WKB theory 
	alone, and showed that they agreed well with the full numerical eigenvalues (Figures \ref{fig:quantumconditionloopg1}$-$\ref{fig:quantumconditionloopg20}).

	\end{enumerate}

\acknowledgments

We thank Pak-Shing Li for a discussion on the self-gravity kernels and Phil Nicholson and Yanqin Wu for helpful discussions. Y.L. acknowledges NSF grant AST-1352369 and NASA grant  NNX14AD21G.

\appendix

\section{Derivation of the Governing Equation}
\label{sec:linearization}
\setcounter{equation}{0}
\subsection{Linear Equations}

We consider  linear perturbations by expanding the fluid variables as $X+X_1$ where the subscript 1 denotes the perturbation. Since the coefficients have no explicit dependence on $t$ and $\varphi$, the equations can be Fourier transformed in terms of wave frequency $\omega$ and azimuthal wavenumber $m$:
\begin{align}
\label{eq:app1_1}
X_1(r,\varphi,t) = {\rm Re}\left[X_{1,m}(r) e^{i(m\varphi-\omega t)}\right],
\end{align}
where $X_{1,m}$ is the $m$-th Fourier component of $X_1$. As we consider only a single $m$, the subscript $m$ is dropped for clarity. We denote $u=u_{r1}$, and $v=u_{\varphi 1}$ as the radial and azimuthal velocity of the linear perturbation, respectively. The linearized continuity and momentum equations now read
\begin{align}
\label{eq:app1_3a}
-i\hat{\omega}\left(\frac{\Sigma_1}{\Sigma}\right) &+ \frac{d u}{dr} + u\frac{d}{dr}\ln(r\Sigma)+\frac{imv}{r} =0 \ , \\
\label{eq:app1_3b}
-i\hat{\omega}u &- 2\Omega v = -\frac{1}{\Sigma}\frac{dP_1}{dr} + \left(\frac{\Sigma_1}{\Sigma}\right)\frac{1}{\Sigma}\frac{dP}{dr}-\frac{d\phi_1}{dr} \ , \\
\label{eq:app1_3c}
-i\hat{\omega}v &+ \frac{\kappa^2}{2\Omega} u = - \frac{imP_1}{r\Sigma}-\frac{im}{r}\phi_1 \ ,
\end{align}
where $\hat{\omega} = \omega - m\Omega$ is the Doppler-shifted frequency, $\Sigma_1$, $P_1$, and $\phi_1$ are the perturbations of surface density, pressure, and self-gravity potential, respectively. The epicyclic frequency $\kappa$ is defined as
\begin{align}
\kappa^2 = \frac{1}{r^3}\frac{d}{dr}\Big(r^4\Omega^2\Big) \ .
\label{eq:kappa}
\end{align}
Next, we eliminate $v$ by using Equations \eqref{eq:app1_3b} and \eqref{eq:app1_3c}, which gives
\begin{align}
\label{eq:app1_4}
D u &= \frac{i\hat{\omega}}{\Sigma}\frac{dP_1}{dr} - i\hat{\omega}\left(\frac{\Sigma_1}{\Sigma}\right)\left(\frac{1}{\Sigma}\frac{dP}{dr}\right)-\frac{2im\Omega}{r}\frac{P_1}{\Sigma} 
+ i\hat{\omega}\frac{d\phi_1}{dr}-\frac{2im\Omega\phi_1}{r} \ ,
\end{align}
where $D=\kappa^2 - (\omega-m\Omega)^2$ is the resonant parameter. To relate pressure and density, we need to specify the equation of state. For adiabatic perturbations, we set the linearized entropy perturbation to be zero, i.e., 
\begin{align}
\label{eq:app1_3d}
-i\hat{\omega}\left(\frac{P_1}{P}-\gamma\frac{\Sigma_1}{\Sigma}\right) + u \left(\frac{1}{P}\frac{dP}{dr}-\frac{\gamma}{\Sigma}\frac{d\Sigma}{dr}\right) =0 \ ,
\end{align}
where $\gamma$ is the 2D adiabatic index. 

Apart from the self-gravity terms (i.e., $\phi_1$), Equations \eqref{eq:app1_3a}, \eqref{eq:app1_4}, and \eqref{eq:app1_3d} can be reduced into a single differential equation for general $m$ \citep[e.g.,][]{1973StAM...52....1F,1979ApJ...233..857G,2014ApJ...782..112T}. Before simplifying these equations further, we address the resonant parameter, which for slow modes ($m=1$ and $|\omega| \ll \Omega$), can be approximated as $D\simeq 2\Omega\left( \omega+{\kappa^2-\Omega^2\over 2\Omega} \right)$. For small $\Omega-\kappa$,
\begin{align}
\label{eq:app1_10}
D \simeq 2\Omega[\omega - (\Omega-\kappa)] \ ,
\end{align}
where, from Equations  \eqref{eq:basic2_1a} and \eqref{eq:kappa}
\begin{align}
\label{eq:app1_11}
\Omega-\kappa &= -\underbrace{
	\frac{1}{2\Omega}\left[\frac{1}{r^2}\frac{d}{dr}\left(\frac{r^2 }{\Sigma}\frac{dP}{dr}\right)\right]}_{\rm pressure}
-\underbrace{\frac{1}{2\Omega}\left[\frac{1}{r^2}\frac{d}{dr}\left(r^2\frac{d\phi_0}{dr}\right)\right]}_{\rm self-gravity} \ ,
\end{align}
with $\phi_0$ given explicitly by Equation \eqref{eq:app1_21} below.
Equation \eqref{eq:app1_11} shows that $\Omega-\kappa$ is indeed small because the pressure
term is  $\sim h^2\Omega$ and the self-gravity term is  $\sim  \mu \Omega$ (Equation \eqref{eq:hmu}), and
so both must be $\lesssim \Omega$ in a stable disk.

\subsection{Eccentric Mode}

To obtain an eccentricity equation, we express the perturbation variables in terms of eccentricity $E$. We define the complex eccentricity as $E = |E| e^{-i\varpi}$ where $|E|$ is the amplitude and $\varpi$ is argument of periapse \citep{2006MNRAS.368.1123G}. Using the formula of an elliptic orbit, the velocity components and the surface density to linear order in $E$ can be expressed as \citep{Pap2002}:
\begin{align}
\label{eq:app1_6a}
u = ir\Omega E \ , \quad v = -\frac{1}{2}r\Omega E \ , \quad \text{and} \quad 	\Sigma_1 = -r\frac{d}{dr}\big(\Sigma E\big) \ ,
\end{align}
where we used the continuity equation  (Equation \eqref{eq:app1_3a}) to relate $\Sigma_1$ to $E$. When evaluating $u$ and $v$, 
we may set $\Omega$ to its Keplerian value, i.e., we may drop the small corrections that are of order $h^2$ and $\mu$.  For adiabatic gas, the pressure perturbation $P_1$ reads
\begin{align}
\label{eq:app1_6c}
P_1 &= - \gamma r P\frac{dE}{dr} - r \frac{dP}{dr} E \ ,
\end{align}
which is obtained by substituting $u$ and $\Sigma_1$ into Equation \eqref{eq:app1_3d}. Substituting the expressions of $u$, $\Sigma_1$, and $P_1$ in terms of $E$ into Equation \eqref{eq:app1_4} and assuming $\hat{\omega} \simeq -\Omega$, we obtain \citep{Pap2002,2006MNRAS.368.1123G,2016MNRAS.458.3221T}
\begin{align}
\label{eq:app1_12a}
2r^3\Omega\Sigma \omega E &= \frac{d}{dr}\left(\gamma r^3 P \frac{dE}{dr}\right) + r^2 \frac{dP}{dr} E - r\Sigma\frac{d}{dr}\left(r^2\frac{d\phi_0}{dr}\right)E - \Sigma \frac{d}{dr}(r^2\phi_1) \ ,
\end{align}
which is Equation \eqref{eq:basic3_2}. Zero Lagrangian pressure perturbation is used for the boundary condition, i.e., $P_1 + rE dP/dr = 0$, which implies $dE/dr=0$ according to Equation \eqref{eq:app1_6c} \citep{2016MNRAS.458.3221T}.

\subsection{Self-Gravity}
\label{sec:app1_selfgravity}

To close the equation, we need to express the self-gravity potential perturbation $\phi_1$ in terms of $E$ and to express $\phi_0$ in terms of $\Sigma$. The Poisson equation for an $m$-mode in a two-dimensional disk with zero-thickness reads
\begin{align}
\label{eq:app1_20}
\left(\frac{\partial^2}{\partial r^2} + \frac{1}{r}\frac{\partial}{\partial r} -\frac{m^2}{r^2} + \frac{\partial^2}{\partial z^2}\right) \phi_m = 4 \pi G \Sigma_m \delta(z) \ ,
\end{align}
where $\delta(z)$ is the Dirac-delta function,  and $\Sigma_m(r)$ and $\phi_m(r,z)$ are the $m$-th Fourier component of the surface density and potential, respectively. As the basic state is axisymmetric, we have $\Sigma=\Sigma_{0}$. In this work, we only consider the potential at the midplane, i.e., $\phi_m(r)=\phi_m(r,z=0)$. The solution to the Poisson equation reads \citep{1992pavi.book.....S}
\begin{align}
\label{eq:app1_21}
\phi_m(r) &= -2\pi G \int K^{(m)}(r,s)\Sigma_m(s) sds  \ ,
\end{align}
where $K^{(m)}(r,s)$ is the Poisson kernel for $m$-th harmonics given by
\begin{align}
\label{eq:app1_22}
K^{(m)}(r,s) = \frac{1}{\pi}\int^\pi_0 \frac{\cos m\varphi}{\sqrt{r^2 + s^2 -2rs\cos\varphi}}d\varphi \ .
\end{align}
Thus, $\phi_0 = -2\pi G \int K^{(0)}(r,s)\Sigma(s) s ds$ and $\phi_1 = -2\pi G \int K^{(1)}(r,s)\Sigma_1(s) s ds$, closing Equation \eqref{eq:app1_12a}. We neglect the indirect potential because it is proportional to $\omega^2$ \citepalias{1990ApJ...358..495S} and is smaller than other terms by $\omega$ for $m=1$ slow modes.

\section{Method of Solutions}
\label{sec:methodofsolution}
\setcounter{equation}{0}

In this section, we describe the method of solution to Equation \eqref{eq:basic3_2}, together with Equations \eqref{eq:app1_21}--\eqref{eq:app1_22} and \eqref{eq:app1_6a} for $\phi_0$ and $\phi_1$. A finite difference method is developed to obtain the eigenfrequencies and eigenfunctions of the boundary-value problem. While the finite difference method is relatively standard, our softening-free implementation of self-gravity is new. Implementation in Appendix \ref{sec:ms_implementation} is general for both softened and unsoftened self-gravity kernels. In Appendix \ref{sec:numericalpoisson}, we describe 
how we treat the singular diagonal components of the gravity kernels without explicit softening. In Appendix \ref{sec:convergence}, we demonstrate the numerical convergence of our implementations. Throughout this paper we do not use softening, except in Appendix \ref{sec:convergence}, for comparison purposes.

\subsection{Implementation of the Boundary Value Problem}
\label{sec:ms_implementation}
We work on a uniform log-spaced grid, with ratio of spacings $\beta=(r_{\rm max}/r_{\rm min})^{1/N}$, where $r_{\rm max}$ ($r_{\rm min}$) is the outer (inner) computational boundary and $N$ is the number of grid points. Variables live at the cell-center and are imagined to be piecewise constant. Equation \eqref{eq:basic3_2} is  discretized and written in the following matrix form:
\begin{align}
\label{eq:ms1_1}
\mathbf{M}{\bm E} = \omega {\bm E} \ ,
\end{align}
where ${\bm E}$ is a column vector of the eccentricity on the radial grid (length $N$) and $\mathbf{M}$ is an $N\times N$ matrix. The  matrix $\mathbf{M}=\mathbf{M}^{(1)}+\mathbf{M}^{(2)}$ is a sum of  $\mathbf{M}^{(1)}$ that contains the derivatives and linear terms in Equation \eqref{eq:basic3_2}, and   $\mathbf{M}^{(2)}$ implements the $\phi_1$ term.

The matrix $\mathbf{M}^{(1)}$ is obtained by discretizing the following part of Equation \eqref{eq:app1_12a},
\begin{align}
\label{eq:ms1_3}
{\gamma P\over 2\Omega \Sigma}
\frac{d^2}{dr^2} + \frac{\gamma}{2r^3\Omega\Sigma}\frac{d}{dr}(r^3 P)\frac{d}{dr} + \left[\frac{1}{2\Omega r\Sigma}\frac{dP}{dr} - \frac{1}{2\Omega r^2}\frac{d}{dr}\left(r^2\frac{d\phi_0}{dr}\right)\right] \ ,
\end{align}
and by replacing the derivatives that act on $E$ with differentiation matrices. Here, a $2^{\rm nd}$-order central finite difference scheme for uniform log-spacing is used and we define $D_{ij}^{(1)}$ and $D_{ij}^{(2)}$ as the differentiation matrices for the first and second derivatives, respectively. Then, the derivatives at the $i$-th radial cell are given by \citep[e.g.,][]{1989ApJ...347..959A}
\begin{align}
\label{eq:ms1_4}
\frac{dE}{d\ln r}\bigg|_i  =\sum_j D_{ij}^{(1)} E_j, \quad \text{and} \quad \frac{d^2 E}{d (\ln r)^2}\bigg|_i  =\sum_j D_{ij}^{(2)} E_j \ .
\end{align} 
For the first and last rows of $\mathbf{M}^{(1)}$, we replace the ghost cells (exterior points) from the stencil by the appropriate interior cells by using the zero Lagrangian pressure perturbation boundary condition $dE/dr=0$ (Appendix \ref{sec:linearization}).

The potential $\phi_0$ of the basic state in $\mathbf{M}^{(1)}$ is computed using Equations \eqref{eq:app1_21}--\eqref{eq:app1_22} in advance, which results in the following discretized form (for general $m$-th Fourier component)
\begin{align}
\label{eq:ms1_5}
\phi_{i}^{(m)} = -2\pi G \sum_j  K_{ij}^{(m)} \Sigma_m(r_j) r_j \Delta r_j \ ,
\end{align}
where $\Delta r_j$ is the width of the cell and $K^{(m)}_{ij} = K^{(m)}(r_i,r_j)$. Note that $K^{(m)}_{ij}$ is singular at $i=j$, an issue we return to in the following subsection.
After $\phi_0$ is obtained, we numerally differentiate it using $2^{\rm nd}$-order finite difference scheme to obtain the last self-gravity term in Equation \eqref{eq:ms1_3}. For the  second matrix $M^{(2)}_{ij}$, we  express
the $\phi_1$ term in Equation \eqref{eq:basic3_2} as
\begin{align}
\label{eq:app1_24}
\frac{d}{dr}(r^2\phi_1) = -2\pi G \int \partial_r \partial_{s} \left[r^2 s^2 K^{(1)}(r,s)\right] \Sigma(s) E(s) ds \ ,
\end{align}
after making use of the continuity equation (Equation \ref{eq:app1_6a}) and integrating by parts. In the above expression, we neglect boundary terms because $\Sigma$ is vanishingly small at the boundaries; but even if $\Sigma$ were finite at the boundaries, the above expression would remain true because of the direct edge contribution that cancels the boundary terms left over by the integration by parts \citepalias[see a discussion in][]{1990ApJ...358..495S}. The matrix $M_{ij}^{(2)}$ is then given by,
\begin{align}
\label{eq:ms1_6}
M^{(2)}_{ij} &= \frac{\pi G}{\Omega_i r_i^3 } P_{ij}\Sigma_j \Delta r_j \ ,
\end{align}
where $\Omega_i=\Omega(r_i)$, $\Sigma_j=\Sigma(r_j)$, and the kernel is $P_{ij} = (\partial_r \partial_{s} [r^2 s^2 K_1(r,s)])_{r_i,r_j}$. Again, the kernel is singular at $i=j$. Here, we do not simplify $P_{ij}$ as a Laplace coefficient \citep[i.e., $b^{(1)}_{3/2}$, as done for an external mass by][]{2010MNRAS.406.2777L}. Instead, we express the kernel in the following symmetric form ($\mathbf{P} = \{P_{ij}\}$):
\begin{align}
\label{eq:ms1_7}
\mathbf{P} = \mathbf{D} \mathbf{Q} \mathbf{D}^T \ ,
\end{align}
where $\mathbf{D} = \{r_i^{-1}D^{(1)}_{ij}\}$ is a differentiation matrix (first derivative), and the superscript $T$ denotes transpose. The elements of $\mathbf{Q}$ are $Q_{ij} = r_i^2 r_j^2 K_1(r_i, r_j)$. The above expression is a pair of matrix-multiply operations and essentially applies the numerical differentiations on $K^{(1)}$, in which one can use the softened kernel $K^{(1)}$ or the unsoftened implementation described below. 

\subsection{Resolving the Singularity in the Diagonal  Terms}
\label{sec:numericalpoisson}

As mentioned above, the diagonals of the kernels, i.e., $K^{(m)}(r,r)$, are singular. There are at least two ways to resolve the issue:
a) use the softened counterpart, as is typically done in the literature (e.g., \citetalias{2001AJ....121.1776T}, \citealt{2008ApJ...678..483B}); b) solve it exactly without softening \citep{1996ApJ...460..855L}, as we present here. 

For $K^{(0)}$, recall that Equation \eqref{eq:ms1_5} originated from Equation \eqref{eq:app1_21}. 
We cast the diagonal term into an integral over the cell assuming the surface density is constant in the cell. Thus, we express the diagonal term in Equation \eqref{eq:ms1_5} as:
\begin{align}
\label{eq:ms1_10}
K^{(m)}_{ii} \rightarrow \frac{1}{r_i\Delta r_i}\int_{\hat{r}_{i}}^{\hat{r}_{i+1}} \sqrt{r_i s}\,K^{(m)}(r_i, s) ds \ ,
\end{align}
where $\hat{r}_i$ denotes the left edge of $i$-th cell. The square root factor is chosen such that the right-hand side is $a/r_i$, where $a$ is a number independent of $i$, so that we need only compute the integral once. The integral can be separated into two pieces ($s<r_i$ and $s>r_i$) and evaluated using the open-ended Newton-Cotes quadrature formula to avoid the singularity at $s=r_i$ \citep[e.g.,][]{1989ApJ...347..959A,1996ApJ...460..855L}.

For the self-gravity perturbation term (Equation \eqref{eq:ms1_6}), we similarly replace the diagonal term of $\mathbf{Q}$ in Equation \eqref{eq:ms1_7} by its cell-averaged value, i.e.,
\begin{align}
\label{eq:ms1_11}
Q_{ii} = \frac{1}{\Delta r_i}\int_{\hat{r}_{i}}^{\hat{r}_{i+1}} r_i^2 s^2 K^{(1)}(r_i, s)\, ds \ ,
\end{align}
which can be evaluated by the same method mentioned above. 

Finally, we note that the non-diagonal terms of the kernel (Equation \eqref{eq:app1_22}) can be expressed as 
\begin{align}
\label{eq:conv1_1}
K_{ij}^{(m)} = \frac{\alpha^{1/2}}{2\sqrt{r_i r_j}}b^{(m)}_{1/2}(\alpha) \ ,
\end{align}
where $\alpha = \min(r_i,r_j)/\max(r_i,r_j) = \beta^{-|i-j|}$ ($\beta$ is the width ratio between adjacent grid cells), and $b^{(m)}_{\rm s}(\alpha)$ is the Laplace coefficient \citep[cf. Equation 6.67 of][]{1999ssd..book.....M}. In practice, we express $b^{(m)}_{\rm s}(\alpha)$ as a series solution involving hypergeometric function $_2 F_1$ and evaluate it numerically. Since $\sqrt{r_i r_j} K^{(m)}_{ij}$ depends only on $|i − j|$ (including $i=j$), it can be evaluated for the first row ($i = 0$) and populated to other rows by shifting the index. 

\subsection{Convergence}
\label{sec:convergence}

\subsubsection{Self-gravity}
\begin{figure*}
	\centering
	\includegraphics[trim={1cm 0.3cm 0 0},clip,width=\textwidth]{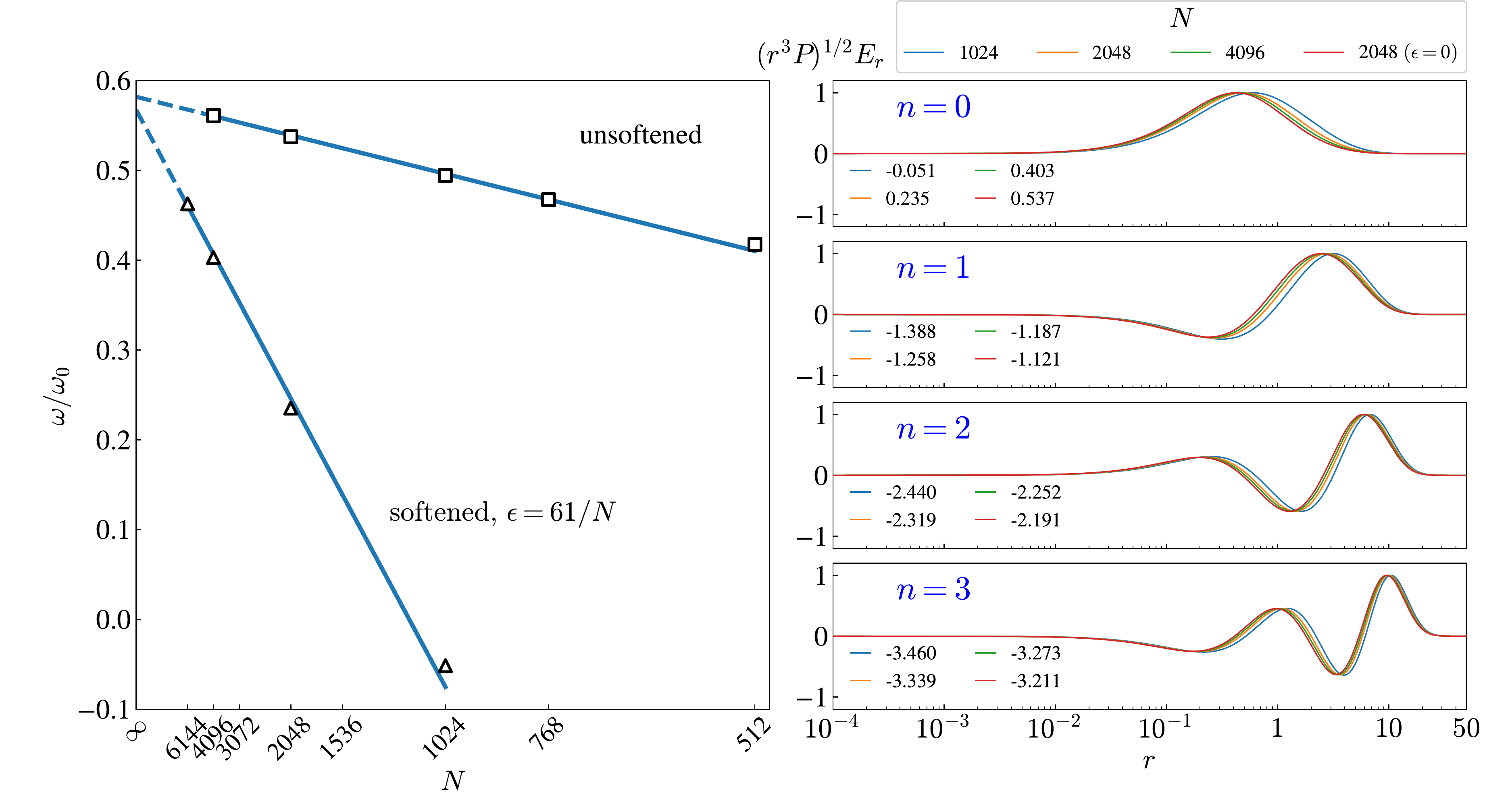}
	\caption{Convergence of eigenfrequencies with $N$ for softened and unsoftened implementations.
	We consider here the fundamental mode in the $g_{\rm max}=1$ fiducial simulation. The blue straight lines are the linear fit to the data, and the dashed portions are the extrapolation. (\emph{Right}) From top to bottom row, we show the eigenfunctions $y=(r^3 P)^{1/2}E_r$ for the first 4 modes. At each row, we compare the three softened solutions ($N=1024$, 2048, and 4096) and the unsoftened solution ($N=2048$). The eigenvalues $\omega/\omega_0$ are shown at the lower-left corners. The eigenfunctions are generally not sensitive to the softening.}
	\label{fig:omega-convergence}
\end{figure*}

To demonstrate convergence of our softening-free method, we calculate the eigenfrequencies of the fundamental mode in our fiducial $g_{\rm max}=1$ calculation (\S \ref{sec:suiteofsixmodels}), for various numbers of grid elements $N$.
Figure \ref{fig:omega-convergence} (squares in left panel) shows the result,  and demonstrates that our unsoftened method converges
towards high $N$.
In this paper
we use $N=2048$, 
which for the case shown gives an eigenfrequency that is accurate to within $\sim 6\%$. The figure shows first-order convergence ($\propto N^{-1}$), as is expected from our numerical method that uses piece-wise constant gravity kernel evaluation \citep[see][for a discussion of convergence]{2016ApJS..224...16W}.

We also examine softened potentials by replacing 
\begin{align}
\label{eq:con1_1}
|\mathbf{r}-\mathbf{r}'| \rightarrow (r^2+r'^2-2rr'\cos{\varphi}+ \epsilon^2 r r')^{1/2} \ ,
\end{align}
in the denominator of the Kernel definition (Equation \eqref{eq:app1_22}) \citep[e.g.,][]{2008ApJ...678..483B}. The triangles in the left panel of Figure demonstrate convergence with increasing $N$ when we fix $\epsilon=61/N$, which corresponds to 
a softening lengthscale ($\epsilon r$) that is larger than the width of a grid element $(\Delta r)$ by a factor of $61/\ln(r_{\rm max}/r_{\rm min})\approx$ 5. We find that 5 grid elements per softening lengthscale is adequate to resolve it; i.e., increasing $N\epsilon$ to more than 61 does not affect the results.

In the right panel, we show the eigenfunctions and eigenfrequencies for the fundamental mode, as well as the first three harmonics,  both with and without softening.
The unsoftened solution at $N=2048$ is shown in red and is located on the left. For the softened solutions, the eigenfunctions shift to the left for increasing $N$ at fixed $N\,\epsilon$. Overall shapes of the eigenfunctions are similar among different cases of $N$ and $\epsilon$.

\subsubsection{Inner Radius}

\begin{figure}[t]
	\centering
	\includegraphics[trim={0 0.3cm 0 0},clip,width=0.5\linewidth]{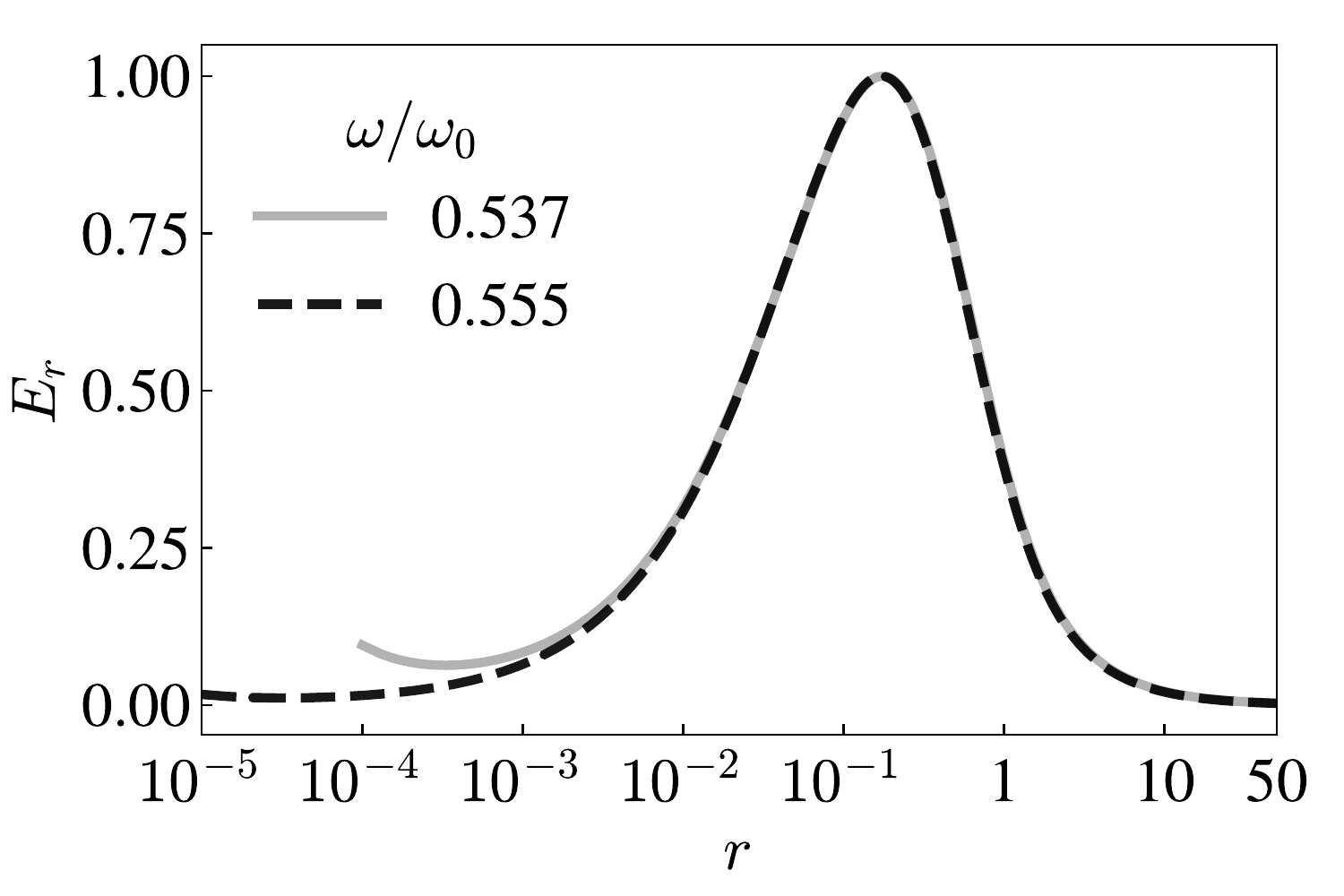}
	\caption{Comparison between the fundamental modes of a $g_{\rm max}=1$ disk with $r_{\rm min}=10^{-4}$ (solid) and $10^{-5}$ (dashed). The eigenfrequencies are similar and the eigenfunctions are almost indistinguishable except near the inner edge.}
	\label{fig:Collection_gmax1}
\end{figure}

Figure \ref{fig:Collection_gmax1} shows the effect of moving the inner boundary location ($r_{\rm min}$). The modes behave very similar except near the inner edge. This is because the inner boundary is located outside the wave cavity (Figure \ref{fig:Collection}) and hence does not affect the mode significantly.

\section{Derivation of WKB2}
\label{sec:derivationofDR2}
\setcounter{equation}{0}

Starting from the linear equation of motion (Equation \eqref{eq:basic3_2}), 
we derive the second-order WKB dispersion relation (Equation \eqref{eq:asymp1_1a}), which contains terms that are smaller than the leading ones by $1/|kr|^2$. The pressure and self-gravity parts of the governing equation  can be treated separately.

\subsection{Pressure}
For the pressure part, we  remove the single derivative term ($\propto dE/dr$) by using an integrating factor, and taking $d^2/dr^2 \rightarrow -k^2$ \citep[cf.][]{2007AN....328..273G}. This procedure automatically yields the
second-order dispersion relation, which we will present the details of in Paper II.
To be specific, we make the following transformation
\begin{align}
\label{eq:app5_1}
y = (r^3 P)^{1/2}E \ .
\end{align}
Physically, $y$ is related to the angular momentum flux.  In particular the linear equation
implies  $dF/dr=0$ where 
$F={\rm const.}\times {\rm Im}\left[ (r^3P) E^* {dE\over dr}\right]={\rm const}\times
{\rm Im}\left(y^*{dy\over dr} \right)$ is the flux, neglecting self-gravity terms. Equation \eqref{eq:basic3_2} now has no single derivative terms of $E$:
\begin{align}
\label{eq:app5_2}
c^2 \left[\frac{d^2}{dr^2} + \frac{2\Omega}{c^2}\left(\omega_p+\omega_g^0-\omega \right)\right] y &= 
-(r^3 P)^{1/2}\frac{1}{r^3}\frac{d}{dr}(r^2 \phi_1) \ ,
\end{align}
where  
\begin{eqnarray}
\label{sec:app5_3}
\omega_p &=& -\frac{c^2}{2\Omega}\left[(r^3 P)^{-1/2}\frac{d^2}{dr^2}(r^3 P)^{1/2}-\frac{1}{\gamma rP}\frac{dP}{dr}\right] \ ,\\
\omega_g^0&=&-{1\over 2\Omega r^2}{d\over dr}\left( r^2{d\phi_0\over dr} \right) \ .
\end{eqnarray}
We may now replace $d^2y/dr^2\rightarrow -k^2y$ into Equation \eqref{eq:app5_2}, 
yielding for the left-hand side:
\begin{align}
\label{eq:app5_5}
\text{L.H.S.} = 2\Omega\left(-{k^2c^2\over 2\Omega} + \omega_p+\omega_g^0-\omega\right)y \ ,
\end{align}
thereby accounting for the pressure terms in Equation \eqref{eq:asymp1_1a}, as well as one part of the $\omega_g$
term in that equation---specifically, the first term in Equation \eqref{eq:omg}. 
It remains to address the right-hand side of Equation  \eqref{eq:app5_2}, as we do in the following.

\subsection{Self-Gravity}

\paragraph{Leading-order WKB}  Using the well-known leading-order solution of the Poisson equation \citep{1964ApJ...140..646L} 
\begin{align}
\label{eq:app5_6b}
\phi_1 = -\frac{2\pi G\Sigma_1}{|k|} \ ,
\end{align}
and $\Sigma_1 = -ikr\Sigma E$ (Equation \eqref{eq:app1_6a}),
the right-hand side of Equation \eqref{eq:app5_2} is $-2\pi G\Sigma|k|y$, which yields the usual leading-order self-gravity term in the dispersion relation.

\paragraph{Second-order WKB} 

To obtain a second-order WKB expression of the right-hand side of Equation \eqref{eq:app5_2}, we take advantage of the following second-order-accurate expression from 
\citet{1979SJAM...36..407B} 
\footnote{For our Equation \eqref{eq:app5_12a}, we 
use Equation (35) in  \citet{1979SJAM...36..407B}, after setting $ik\rightarrow d/dr$ in the first term of that equation, and  $m=1$. Note that the
 ``three orders in the asymptotic expansion" accuracy described in \citet{1979SJAM...36..407B} means second-order-accurate in our definition. }

\begin{align}
\label{eq:app5_12a}
\frac{1}{\sqrt{r}}\frac{d}{dr}(\sqrt{r}\phi_1) + \frac{3i}{8kr^2}\phi_1 = -i{|k|\over k} 2\pi G \Sigma_1  \ .
\end{align}
Therefore, on the right-hand side of Equation \eqref{eq:app5_2} we may substitute
\begin{align}
\label{eq:app5_12b}
\frac{1}{r^2}\frac{d}{dr}(r^2 \phi_1) = \frac{1}{\sqrt{r}}\frac{d}{dr}(\sqrt{r} \phi_1) + \frac{3}{2r}\phi_1
= -i{|k|\over k} 2\pi G \Sigma_1+\frac{3}{2r} \left(1 - \frac{i}{4kr}\right) \phi_1 \ .
\end{align}
Since to the leading-order, $\phi_1$ is  $1/|kr|$ smaller than $\Sigma_1$, the tail $\phi_1$ term in Equation \eqref{eq:app5_12b} is at least $1/|kr|$ smaller than the first term. Thus, it suffices to replace the tail $\phi_1$ term with a solution that is accurate to first order. To do so, we use  Equations (33) and (34) of \citet{1979SJAM...36..407B}, after expanding their latter equation to first order in $k$:
\begin{align}
\phi_1 = - \left[\frac{1}{(k^2r^2+9/4)^{1/2}}
+\frac{i kr}{2\left(k^2r^2+9/4\right)^{3/2}}+\mathcal{O}\left(\frac{1}{|kr|^2}\right)\right]2\pi G\Sigma_1 \ .
\end{align}
(Note that we actually expand in $kr/(k^2r^2+9/4)$ rather than $1/(kr)$ so that we can evaluate our final dispersion relation at Lindblad resonances, which are at $k=0$. Of course, the WKB approximation breaks down at $|kr|\lesssim 1$; but with this expansion, we find our final expression at $k=0$ is similar enough to that at $|kr|\sim 1$.)

Equation \eqref{eq:app5_12b} becomes
\begin{align}
\label{eq:app5_12d}
\frac{1}{r^2}\frac{d}{dr}(r^2 \phi_1) &= -\left\lbrace{|k|\over k} i + \frac{3}{2\sqrt{k^2 r^2 + 9/4}}\left[1 - \frac{i}{4kr} + \frac{ikr}{2(k^2r^2+9/4)}\right]\right\rbrace2\pi G\Sigma_1 \ .
\end{align}
After substituting the WKB expression for $\Sigma_1$ that includes the gradient of the basic state,
\begin{align}
\label{eq:app5_12e}
\Sigma_1 = -(r^3 P)^{-1/2} \Sigma \left\lbrace\frac{d}{d\ln r} \ln \left[\frac{\Sigma}{(r^3 P)^{1/2}}\right]+ikr\right\rbrace y \ ,
\end{align}
(cf., Equations \eqref{eq:app1_6a} and \eqref{eq:app5_1}), 
the right-hand side of Equation \eqref{eq:app5_2} reads
\begin{equation}
{\rm R.H.S.} = 2\Omega \left\lbrace - \frac{\pi G\Sigma}{\Omega} |k| + \left(\frac{\pi G \Sigma}{r\Omega}\right)\frac{3}{2\sqrt{k^2 r^2 + 9/4}} \left[\frac{d}{d\ln r}\ln\left(\frac{\Sigma}{(r^3 P)^{1/2}}\right)-\frac{1}{4}\right]+ {\rm imaginary\ terms}    \right\rbrace y \ ,
\end{equation}
where we suppress the imaginary terms ($\propto i$). 
Equating with Equation \eqref{eq:app5_5}, we recover WKB2 (Equation \ref{eq:asymp1_1a}) after dropping the imaginary terms. We note that the second term in braces (i.e., the second term of $\omega_g$ in Equation \eqref{eq:omg}) is caused by the fact that $\phi_1$ and $\Sigma_1$ are no longer perfectly in-phase (cf. Equation \eqref{eq:app5_12d}). This happens when the spiral waves become more open. 

\section{Quantum Condition}
\label{sec:q}
To obtain the quantum condition for slow modes (Equation \eqref{eq:quantumcondition0}), one needs the sum of all phase changes
at LRs and QBs.  In Appendix \ref{sec:noteonlrqb}, we calculate the phase change at LRs explicitly, because it has been treated
incorrectly in the literature.  Then in Appendix \ref{sec:totalphasechange}, we show that the total phase change over a cycle is always $\pi$ for  slow modes.

\subsection{Phase Change at LRs}
\label{sec:noteonlrqb}
\setcounter{equation}{0}

Following  \citetalias{1990ApJ...358..495S}, 
we first decompose the equation of motion into two separate ones---one for the leading component, 
and one for the trailing.
We restore the equation of motion from the WKB2 dispersion relation (Equation \ref{eq:asymp1_1a}) by replacing $k\rightarrow  -id/dr$, and then ``operating''   the dispersion relation   on $y=(r^3P)^{1/2}E$ (reversing the steps in Appendix \ref{sec:derivationofDR2}).  We then decompose $y$ into trailing($+$) and leading($-$) components: 
$y=y_++y_-$, in which case the $|k|\rightarrow  \mp id/dr$, and
the following equation of motion results:
\begin{align}
L_+ y_+ + L_- y_- = 0 \ ,
\end{align}
where
\begin{eqnarray}
\label{eq:eq10_4}
L_\pm &=& \pm i 2\pi G \Sigma \frac{d}{dr} + 2\Omega (\omega-\omega_{\rm LR}) \ ,\\ 
\omega_{\rm LR}&=&\omega_p+\omega_{g}\vert_{k=0} \ ,
\end{eqnarray}
where we have dropped the $c^2k^2$ pressure term from the dispersion relation
because we are focusing on  LRs, where the long-wave dominates. We split the equation into trailing/leading waves separately, by introducing a complex constant $C$:
\begin{align}
\label{eq:eq10_3b}
L_\pm y_\pm = \pm C \ .
\end{align}
This allows us to solve $y_\pm $ and compare the phase before and after the reflection.

Next, we follow the standard procedure by Taylor expanding  the second term of $L_\pm$ near LR and then matching the asymptotic solutions. Equation \eqref{eq:eq10_3b} can be rewritten as
\begin{align}
\label{eq:eq10_7}
\left(\frac{d}{d\zeta} \mp i s_\mathscr{D} \zeta\right) y_\pm = 1 \ ,
\end{align}
where we define
\begin{align}
\zeta = \left(\frac{|\mathscr{D}|}{2\pi G\Sigma r}\right)^{1/2}_L(r-r_L) \ ,\quad \text{and} \quad \mathscr{D} \equiv -2\left(r\Omega\frac{d\omega_{\rm LR}}{dr}\right)_L \ ,
\end{align}
and $s_\mathscr{D}={\rm sgn}(\mathscr{D})=-{\rm sgn}(d\omega_{\rm LR}/dr)$. Here, the subscript $L$ denotes the quantity to be evaluated at the LR. We note that our derivation differs from \citetalias{1990ApJ...358..495S} in the following ways: i) We keep the sign of $\mathscr{D}$ because $\mathscr{D}$ can be positive or negative for slow modes. This is not considered by \citetalias{1990ApJ...358..495S} as their modes are fast and outside corotation; ii) Unlike the forced treatment in \citetalias{1990ApJ...358..495S} (their Equation (58)), we are free to set the wave amplitude on the right-hand side of Equation \eqref{eq:eq10_7} to unity for free linear waves. 
Integrating Equation \eqref{eq:eq10_7} gives the solution for long trailing/leading waves near the LRs:
\begin{align}
y_\pm \sim \exp \left[\pm s_\mathscr{D}\left(\frac{i\zeta^2}{2}-\frac{i\pi}{4}\right)\right] \ .
\end{align}
The phase factor $\pi/4$ appears as we require the solution to vanish sufficiently far away from the LR in the evanescent region. To obtain the phase change, we need to identify the incoming/outgoing waves. This can be done by computing the phase integral $\int^r k dr = s_\mathscr{D} \zeta^2/2$ for the long branch and matching the asymptotic solutions. Near a LR with $\mathscr{D}>0$, the phase integral increases with $\zeta$ (distance from LRs) and so an incoming leading wave ($y_-$) reflects as a trailing wave ($y_+$) (see the DRM in Figure \ref{fig:illustration}). This gives a $-\pi/2$ phase-shift. On the other hand, a trailing wave reflects into
a leading wave when $\mathscr{D}<0$. Because the incoming/outgoing wave has the same $\pm {\rm sgn}(\mathscr{D})$, the phase-shift is the same. In summary, the phase-shift associated with a LR reads
\begin{align}
\Delta \Phi_{\rm LR} = -\frac{\pi}{4}-\left(\frac{\pi}{4}\right) = -\frac{\pi}{2} \ .
\end{align}
Therefore, $\Delta \Phi_{\rm LR}$ is independent of the sign of $d\omega_{\rm LR}/dr|_L$ and it is the \emph{same} for both ``inner" and ``outer" LRs for slow modes. 

\subsection{Total Phase Change}
\label{sec:totalphasechange}

The total phase change $\Delta \Phi$ (cf. Equation \eqref{eq:quantumcondition0}) is the sum of contributions from individual turning points. The phase change at these locations are
\begin{align}
\label{eq:asymptotic_phaseshift}
\Delta \Phi_{\rm LR} = -\frac{\pi}{2} \quad \text{and} \quad \Delta \Phi_{\rm QB} = +\frac{\pi}{2} \ .
\end{align}
For clarity, we do not include the derivation of $\Delta \Phi_{\rm QB}$ since it is essentially the same as \citetalias{1990ApJ...358..495S} after accounting for a sign difference: we flipped the sign \emph{twice} from their expression because of the opposite sign convention of $k$ and the opposite side of corotation. 

Therefore, for an $s$ mode, there are two QBs and hence $\Delta \Phi = \pi$. 
For $e$-$p$ and $e$-$pg$ modes, which are symmetric in $K$ {on a DRM}, an LR is always accompanied by two QBs. This makes the phase change also $\pi$. In conclusion, $\Delta \Phi = \pi$ for all normal modes.

Although we do not find \citetalias{2001AJ....121.1776T}'s $g$ modes in this paper, they can exist in narrow rings. In that case, the correct quantum condition is 
\begin{align}
\int^{r_+}_{r_-} k dr = \left(n-\frac{1}{2}\right)\pi \ , \quad n=1,2,3,\,\cdots
\end{align}
where $k$ is taken at the long wave branch and $r_\pm$ are the two LRs. This condition differs from \citetalias{2001AJ....121.1776T} because of the incorrect treatment of $\Delta \Phi_{\rm LR}$ there.

\bibliography{disk2,spiraldensitywaves,pressure,eccentricgeneral}

\begin{thebibliography}{}
\expandafter\ifx\csname natexlab\endcsname\relax\def\natexlab#1{#1}\fi
\providecommand{\url}[1]{\href{#1}{#1}}

\bibitem[{{Adams} {et~al.}(1989){Adams}, {Ruden}, \&
  {Shu}}]{1989ApJ...347..959A}
{Adams}, F.~C., {Ruden}, S.~P., \& {Shu}, F.~H. 1989, \apj, 347, 959

\bibitem[{{Ataiee} {et~al.}(2013){Ataiee}, {Pinilla}, {Zsom}, {Dullemond},
  {Dominik}, \& {Ghanbari}}]{2013A&A...553L...3A}
{Ataiee}, S., {Pinilla}, P., {Zsom}, A., {et~al.} 2013, \aap, 553, L3

\bibitem[{{Barker} \& {Ogilvie}(2014)}]{2014MNRAS.445.2637B}
{Barker}, A.~J., \& {Ogilvie}, G.~I. 2014, \mnras, 445, 2637

\bibitem[{{Barker} \& {Ogilvie}(2016)}]{2016MNRAS.458.3739B}
---. 2016, \mnras, 458, 3739

\bibitem[{{Baruteau} \& {Masset}(2008)}]{2008ApJ...678..483B}
{Baruteau}, C., \& {Masset}, F. 2008, \apj, 678, 483

\bibitem[{{Baruteau} \& {Zhu}(2016)}]{2016MNRAS.458.3927B}
{Baruteau}, C., \& {Zhu}, Z. 2016, \mnras, 458, 3927

\bibitem[{{Bertin}(2014)}]{2014dyga.book.....B}
{Bertin}, G. 2014, {Dynamics of Galaxies} (Cambridge, UK: Cambridge University
  Press)

\bibitem[{{Bertin} {et~al.}(1989){Bertin}, {Lin}, {Lowe}, \&
  {Thurstans}}]{1989ApJ...338..104B}
{Bertin}, G., {Lin}, C.~C., {Lowe}, S.~A., \& {Thurstans}, R.~P. 1989, \apj,
  338, 104

\bibitem[{{Bertin} \& {Mark}(1979)}]{1979SJAM...36..407B}
{Bertin}, G., \& {Mark}, J.~W.-K. 1979, SIAM Journal of Applied Mathematics,
  36, 407

\bibitem[{{Binney} \& {Tremaine}(2008)}]{2008gady.book.....B}
{Binney}, J., \& {Tremaine}, S. 2008, {Galactic Dynamics: Second Edition}
  (Princeton, NJ: Princeton University Press)

\bibitem[{{Chan} {et~al.}(2018){Chan}, {Krolik}, \&
  {Piran}}]{2018ApJ...856...12C}
{Chan}, C.-H., {Krolik}, J.~H., \& {Piran}, T. 2018, \apj, 856, 12

\bibitem[{{Dong} {et~al.}(2018){Dong}, {Liu}, {Eisner}, {Andrews}, {Fung},
  {Zhu}, {Chiang}, {Hashimoto}, {Liu}, {Casassus}, {Esposito}, {Hasegawa},
  {Muto}, {Pavlyuchenkov}, {Wilner}, {Akiyama}, {Tamura}, \&
  {Wisniewski}}]{2018ApJ...860..124D}
{Dong}, R., {Liu}, S.-y., {Eisner}, J., {et~al.} 2018, \apj, 860, 124

\bibitem[{{Esposito} {et~al.}(1983){Esposito}, {Borderies}, {Goldreich},
  {Cuzzi}, {Holberg}, {Lane}, {Pomphrey}, {Terrile}, {Lissauer}, {Marouf}, \&
  {Tyler}}]{1983Sci...222...57E}
{Esposito}, L.~W., {Borderies}, N., {Goldreich}, P., {et~al.} 1983, Science,
  222, 57

\bibitem[{{Feldman} \& {Lin}(1973)}]{1973StAM...52....1F}
{Feldman}, S.~I., \& {Lin}, C.~C. 1973, Studies in Applied Mathematics, 52, 1

\bibitem[{{French} {et~al.}(2016){French}, {Nicholson}, {Hedman}, {Hahn},
  {McGhee-French}, {Colwell}, {Marouf}, \& {Rappaport}}]{2016Icar..279...62F}
{French}, R.~G., {Nicholson}, P.~D., {Hedman}, M.~M., {et~al.} 2016, \icarus,
  279, 62

\bibitem[{{Gammie}(2001)}]{2001ApJ...553..174G}
{Gammie}, C.~F. 2001, \apj, 553, 174

\bibitem[{{Goldreich} \& {Tremaine}(1978)}]{1978ApJ...222..850G}
{Goldreich}, P., \& {Tremaine}, S. 1978, \apj, 222, 850

\bibitem[{{Goldreich} \& {Tremaine}(1979)}]{1979ApJ...233..857G}
---. 1979, \apj, 233, 857

\bibitem[{{Goodchild} \& {Ogilvie}(2006)}]{2006MNRAS.368.1123G}
{Goodchild}, S., \& {Ogilvie}, G. 2006, \mnras, 368, 1123

\bibitem[{{Gough}(2007)}]{2007AN....328..273G}
{Gough}, D.~O. 2007, Astronomische Nachrichten, 328, 273

\bibitem[{{Hahn}(2003)}]{2003ApJ...595..531H}
{Hahn}, J.~M. 2003, \apj, 595, 531

\bibitem[{{Hsieh} \& {Gu}(2012)}]{2012ApJ...760..119H}
{Hsieh}, H.-F., \& {Gu}, P.-G. 2012, \apj, 760, 119

\bibitem[{{Lau} \& {Bertin}(1978)}]{1978ApJ...226..508L}
{Lau}, Y.~Y., \& {Bertin}, G. 1978, \apj, 226, 508

\bibitem[{{Lauer} {et~al.}(1993){Lauer}, {Faber}, {Groth}, {Shaya}, {Campbell},
  {Code}, {Currie}, {Baum}, {Ewald}, {Hester}, {Holtzman}, {Kristian}, {Light},
  {Ligynds}, {O'Neil}, \& {Westphal}}]{1993AJ....106.1436L}
{Lauer}, T.~R., {Faber}, S.~M., {Groth}, E.~J., {et~al.} 1993, \aj, 106, 1436

\bibitem[{{Laughlin} \& {Korchagin}(1996)}]{1996ApJ...460..855L}
{Laughlin}, G., \& {Korchagin}, V. 1996, \apj, 460, 855

\bibitem[{{Lee} \& {Goodman}(1999)}]{1999MNRAS.308..984L}
{Lee}, E., \& {Goodman}, J. 1999, \mnras, 308, 984

\bibitem[{{Lin} \& {Shu}(1964)}]{1964ApJ...140..646L}
{Lin}, C.~C., \& {Shu}, F.~H. 1964, \apj, 140, 646

\bibitem[{{Lin} \& {Shu}(1966)}]{1966PNAS...55..229L}
---. 1966, Proceedings of the National Academy of Science, 55, 229

\bibitem[{{Lin} {et~al.}(1969){Lin}, {Yuan}, \& {Shu}}]{1969ApJ...155..721L}
{Lin}, C.~C., {Yuan}, C., \& {Shu}, F.~H. 1969, \apj, 155, 721

\bibitem[{{Lin}(2015)}]{2015MNRAS.448.3806L}
{Lin}, M.-K. 2015, \mnras, 448, 3806

\bibitem[{{Lubow}(2010)}]{2010MNRAS.406.2777L}
{Lubow}, S.~H. 2010, \mnras, 406, 2777

\bibitem[{{Lynden-Bell} \& {Kalnajs}(1972)}]{1972MNRAS.157....1L}
{Lynden-Bell}, D., \& {Kalnajs}, A.~J. 1972, \mnras, 157, 1

\bibitem[{{Lynden-Bell} \& {Pringle}(1974)}]{1974MNRAS.168..603L}
{Lynden-Bell}, D., \& {Pringle}, J.~E. 1974, \mnras, 168, 603

\bibitem[{{Mark}(1976)}]{1976ApJ...205..363M}
{Mark}, J.~W.~K. 1976, \apj, 205, 363

\bibitem[{{Murray} \& {Dermott}(1999)}]{1999ssd..book.....M}
{Murray}, C.~D., \& {Dermott}, S.~F. 1999, {Solar system dynamics} (Cambridge,
  UK: Cambridge University Press)

\bibitem[{{Nicholson} {et~al.}(2014){Nicholson}, {French}, {McGhee-French},
  {Hedman}, {Marouf}, {Colwell}, {Lonergan}, \&
  {Sepersky}}]{2014Icar..241..373N}
{Nicholson}, P.~D., {French}, R.~G., {McGhee-French}, C.~A., {et~al.} 2014,
  \icarus, 241, 373

\bibitem[{{Nicholson} {et~al.}(1978){Nicholson}, {Persson}, {Matthews},
  {Goldreich}, \& {Neugebauer}}]{1978AJ.....83.1240N}
{Nicholson}, P.~D., {Persson}, S.~E., {Matthews}, K., {Goldreich}, P., \&
  {Neugebauer}, G. 1978, \aj, 83, 1240

\bibitem[{{Ogilvie}(2001)}]{2001MNRAS.325..231O}
{Ogilvie}, G.~I. 2001, \mnras, 325, 231

\bibitem[{{Ogilvie}(2008)}]{2008MNRAS.388.1372O}
---. 2008, \mnras, 388, 1372

\bibitem[{{Ogilvie}(2018)}]{2018MNRAS.477.1744O}
---. 2018, \mnras, 477, 1744

\bibitem[{{Ogilvie} \& {Barker}(2014)}]{2014MNRAS.445.2621O}
{Ogilvie}, G.~I., \& {Barker}, A.~J. 2014, \mnras, 445, 2621

\bibitem[{{Pan} {et~al.}(2016){Pan}, {Nesvold}, \&
  {Kuchner}}]{2016ApJ...832...81P}
{Pan}, M., {Nesvold}, E.~R., \& {Kuchner}, M.~J. 2016, \apj, 832, 81

\bibitem[{{Papaloizou}(2002)}]{Pap2002}
{Papaloizou}, J.~C.~B. 2002, \aap, 388, 615

\bibitem[{{Pinte} {et~al.}(2018){Pinte}, {Price}, {M{\'e}nard}, {Duch{\^e}ne},
  {Dent}, {Hill}, {de Gregorio-Monsalvo}, {Hales}, \&
  {Mentiplay}}]{2018ApJ...860L..13P}
{Pinte}, C., {Price}, D.~J., {M{\'e}nard}, F., {et~al.} 2018, \apjl, 860, L13

\bibitem[{{Saini} {et~al.}(2009){Saini}, {Gulati}, \&
  {Sridhar}}]{2009MNRAS.400.2090S}
{Saini}, T.~D., {Gulati}, M., \& {Sridhar}, S. 2009, \mnras, 400, 2090

\bibitem[{{Shu}(1970)}]{1970ApJ...160...99S}
{Shu}, F.~H. 1970, \apj, 160, 99

\bibitem[{{Shu}(1992)}]{1992pavi.book.....S}
---. 1992, {The physics of astrophysics. Volume II: Gas dynamics.} (Mill Valey,
  CA: Univ. Sci. Books)

\bibitem[{{Shu} {et~al.}(1990){Shu}, {Tremaine}, {Adams}, \&
  {Ruden}}]{1990ApJ...358..495S}
{Shu}, F.~H., {Tremaine}, S., {Adams}, F.~C., \& {Ruden}, S.~P. 1990, \apj,
  358, 495

\bibitem[{{Sridhar} {et~al.}(1999){Sridhar}, {Syer}, \&
  {Touma}}]{1999ASPC..160..307S}
{Sridhar}, S., {Syer}, D., \& {Touma}, J. 1999, in Astronomical Society of the
  Pacific Conference Series, Vol. 160, Astrophysical Discs - an EC Summer
  School, ed. J.~A. {Sellwood} \& J.~{Goodman}, 307

\bibitem[{{Tang} {et~al.}(2017){Tang}, {Guilloteau}, {Dutrey}, {Muto}, {Shen},
  {Gu}, {Inutsuka}, {Momose}, {Pietu}, {Fukagawa}, {Chapillon}, {Ho}, {di
  Folco}, {Corder}, {Ohashi}, \& {Hashimoto}}]{2017ApJ...840...32T}
{Tang}, Y.-W., {Guilloteau}, S., {Dutrey}, A., {et~al.} 2017, \apj, 840, 32

\bibitem[{{Teyssandier} \& {Ogilvie}(2016)}]{2016MNRAS.458.3221T}
{Teyssandier}, J., \& {Ogilvie}, G.~I. 2016, \mnras, 458, 3221

\bibitem[{{Toomre}(1969)}]{1969ApJ...158..899T}
{Toomre}, A. 1969, \apj, 158, 899

\bibitem[{{Toomre}(1981)}]{1981seng.proc..111T}
{Toomre}, A. 1981, in Structure and Evolution of Normal Galaxies, ed. S.~M.
  {Fall} \& D.~{Lynden-Bell}, 111--136

\bibitem[{{Tremaine}(1995)}]{1995AJ....110..628T}
{Tremaine}, S. 1995, \aj, 110, 628

\bibitem[{{Tremaine}(2001)}]{2001AJ....121.1776T}
---. 2001, \aj, 121, 1776

\bibitem[{{Tsang}(2014)}]{2014ApJ...782..112T}
{Tsang}, D. 2014, \apj, 782, 112

\bibitem[{{van der Marel} {et~al.}(2013){van der Marel}, {van Dishoeck},
  {Bruderer}, {Birnstiel}, {Pinilla}, {Dullemond}, {van Kempen}, {Schmalzl},
  {Brown}, {Herczeg}, {Mathews}, \& {Geers}}]{2013Sci...340.1199V}
{van der Marel}, N., {van Dishoeck}, E.~F., {Bruderer}, S., {et~al.} 2013,
  Science, 340, 1199

\bibitem[{{van der Plas} {et~al.}(2017){van der Plas}, {M{\'e}nard}, {Canovas},
  {Avenhaus}, {Casassus}, {Pinte}, {Caceres}, \& {Cieza}}]{2017A&A...607A..55V}
{van der Plas}, G., {M{\'e}nard}, F., {Canovas}, H., {et~al.} 2017, \aap, 607,
  A55

\bibitem[{{Wang} {et~al.}(2016){Wang}, {Taam}, \& {Yen}}]{2016ApJS..224...16W}
{Wang}, H.-H., {Taam}, R.~E., \& {Yen}, D.~C.~C. 2016, \apjs, 224, 16

\bibitem[{{Zang}(1976)}]{1976PhDT........26Z}
{Zang}, T.~A. 1976, PhD thesis, , Massachussetts Institute of Technology,
  Cambrigde, MA

\bibitem[{{Zhu} \& {Baruteau}(2016)}]{2016MNRAS.458.3918Z}
{Zhu}, Z., \& {Baruteau}, C. 2016, \mnras, 458, 3918

\end{thebibliography}
\end{document}